\documentclass[aps,prd,twocolumn,nofootinbib,superscriptaddress]{revtex4-2}

\usepackage{amsmath,amssymb,mathtools,bm,mathrsfs}
\usepackage{booktabs}
\usepackage{array}
\usepackage{microtype}
\usepackage{xcolor}
\usepackage[colorlinks=true,linkcolor=blue,citecolor=blue,urlcolor=blue]{hyperref}

\newcommand{\mpl}{\ell_{\rm P}}
\newcommand{\Tr}{\operatorname{Tr}}
\newcommand{\TrF}{\operatorname{Tr}_{F}}
\newcommand{\cG}{\mathcal{G}}
\newcommand{\cR}{\mathcal{R}}
\newcommand{\cE}{\mathcal{E}}
\newcommand{\dd}{\mathrm{d}}
\newcommand{\ii}{\mathrm{i}}

\newcommand{\order}{\mathcal{O}}

\begin{document}

\title{Vacuum Gravity from Entropy: Stability, Spectra, and Exact Waves}

\author{David S. Pereira}
\email{djpereira@ciencias.ulisboa.pt}
	\affiliation{%
		Departamento de F\'{i}sica, Faculdade de Ci\^{e}ncias da Universidade de Lisboa, Campo Grande, Edif\'{\i}cio C8, P-1749-016 Lisbon, Portugal}
	\affiliation{Instituto de Astrof\'{\i}sica e Ci\^{e}ncias do Espa\c{c}o, Faculdade de
		Ci\^{e}ncias da Universidade de Lisboa, Campo Grande, Edif\'{\i}cio C8,
		P-1749-016 Lisbon, Portugal;\\
	}%
\date{\today}

\begin{abstract} 
We analyze the vacuum dynamics of Gravity from Entropy, including its algebraically constrained $G$-field formulation. Evaluating the curvature traces over zero-, one-, and two-form sectors, we show that the complete Minkowski Hessian is exactly that of the quadratic-gravity action $A R+B R_{\mu\nu}R^{\mu\nu}$, with $A=3\beta/\ell_{\rm P}^{4}$ and $B=5\beta^{2}/(2\ell_{\rm P}^{4})$. For diagonalizable curvature blocks, the same action reduces to a sum over eigenvalue logarithms and reproduces these coefficients exactly. A strict diagonal-curvature restriction on the perturbations is instead only a reduced subsector and excludes non-diagonalizable type-N wave curvatures. Linearizing the $G$-field equations and subsequently imposing the algebraic vacuum constraint reproduces the same reduced metric equation and covariant Minkowski Hessian. The spectrum contains the massless graviton, a scalar with $m_{0}^{2}=3/(5\beta)$, and an opposite-residue spin-2 branch with $m_{2}^{2}=-6/(5\beta)=-2m_{0}^{2}$. For the foundational choice $\beta>0$, conventional Einstein normalization therefore implies a tachyonic spin-2 instability. We also show that every four-dimensional Ricci-flat metric solves the local bulk equations through quadratic curvature order, while square-zero Ricci-flat pp-waves are exact local vacuum solutions of the analytic metric-only logarithmic branch. On the isolated massless transverse-traceless eigenspace, the quadratic translation current has the standard general-relativistic normalization. 
\end{abstract}
\maketitle

\section{Introduction}
\label{sec:introduction}
Extensions of General Relativity (GR) constitute a major and extensively developed research programme in contemporary gravitational physics \cite{Clifton:2011jh,Capozziello:2011et,CANTATA:2021asi}. Their study is motivated by unresolved problems spanning cosmology and fundamental theory, including the origin of cosmic acceleration, the physical nature of dark matter and dark energy, the dynamics of the early Universe, and the reconciliation of gravity with quantum physics \cite{Copeland:2006wr,Nojiri:2010wj,Joyce:2014kja,Donoghue:1994dn,Burgess:2003jk}. At the same time, increasingly precise Solar-System, binary-pulsar, cosmological, black-hole, and gravitational-wave observations now probe the gravitational interaction across widely separated length, curvature, and energy scales \cite{Will:2014kxa,Berti:2015itd,Koyama:2015vza,Ishak:2018his,LIGOScientific:2016lio}. These theoretical and observational developments have led to a broad range of modified-gravity frameworks, which must be assessed not only for agreement with existing data, but also for mathematical consistency, absence of pathological degrees of freedom, recovery of the weak-field limit, viable cosmological evolution, and acceptable strong-field phenomenology \cite{Sotiriou:2008rp,DeFelice:2010aj,Joyce:2014kja,CANTATA:2021asi,CosmoVerseNetwork:2025alb}. 

Within this broader programme of constructing and testing theoretically consistent extensions of GR, Gravity from Entropy (GfE) offers a distinct operator-based approach in which the gravitational action is inspired by quantum relative entropy and formulated on the direct sum of zero-, one-, and two-forms \cite{Bianconi:2024aju}. The spacetime metric defines a reference metric on this form space, while curvature and matter modify a second operator that enters a matrix logarithm. In vacuum, the curvature operator is block diagonal, with blocks given by the Ricci scalar, the mixed Ricci tensor, and the Riemann endomorphism acting on two-forms. Within the Taylor domain of the logarithm, the action expands into an infinite sequence of curvature invariants, whose relative combinations are fixed by this direct-sum structure. The numerical value of the trace-log depends only on the eigenvalues of the curvature operator, counted with algebraic multiplicity, and is therefore insensitive to its nilpotent Jordan part. The full tensorial structure nevertheless re-enters through variations of the matrix function and through the auxiliary $G$-field. This theory has been explored in the context of black holes~\cite{Bianconi:2025rnd,Thattarampilly:2026wsw}, inflation~\cite{Thattarampilly:2025krv}, and cosmological thermodynamics~\cite{Bianconi:2025awa}.

The foundational construction defines general flattened operators on the spaces of independent forms and does not require the Ricci or Riemann endomorphisms to be diagonal. Some symmetry-reduced applications subsequently restrict to backgrounds for which the mixed Ricci tensor and the flattened $6\times6$ Riemann matrix are diagonal, thereby converting the trace-log into a sum of scalar logarithms. We distinguish carefully between three operations: choosing an eigenbasis for a diagonalizable operator, substituting a diagonal background into covariantly derived equations, and imposing a fixed diagonal-curvature ansatz before variation. The first two preserve the covariant theory, whereas the third defines a reduced variational problem and is equivalent to the full equations only when the discarded off-diagonal equations vanish identically.

The weak-field content of GfE cannot be read off by simply comparing the action with a familiar higher-derivative model. We must first evaluate the curvature traces over the three form sectors and fix the normalization of the two-form block. This is essential because different conventions for the Riemann endomorphism on antisymmetric tensors differ by factors of two and lead to different quadratic coefficients, pole masses, and residue assignments.

In this work we perform that reduction covariantly, without assuming diagonalizability of the curvature blocks, and show that the complete Minkowski Hessian coincides with that of a specific one-parameter quadratic-gravity theory. We then evaluate the diagonalizable subcase explicitly and show that it gives the same curvature invariants, quadratic coefficients, and pole locations. By contrast, requiring all curvature perturbations to remain diagonal would retain only a restricted subsector of the Hessian. The same covariant linearized equations are obtained directly from the constrained $G$-field formulation, since in vacuum the $G$-field is algebraically determined by the metric curvature. The resulting spectrum contains the massless graviton, a non-tachyonic scalar for the foundational sign of the coupling, and an additional spin-2 branch with opposite residue and negative mass squared. The two nonzero masses are not independent but obey a fixed relation imposed by the GfE direct-sum structure.

The spin decomposition and residue analysis are standard results of four-dimensional quadratic gravity \cite{Rivers:1964nfl,Barnes:1965ylk,VanNieuwenhuizen:1973fi,Stelle:1976gc,Stelle:1977ry,Whitt:1984pd,Hindawi:1995an,Salvio:2018crh}. The GfE-specific result is the coefficient locking that fixes the relative scalar and spin-2 sectors. For the foundational domain $\beta>0$, conventional Einstein normalization and a non-tachyonic scalar necessarily accompany a tachyonic opposite-residue spin-2 branch. Reversing the sign of $\beta$ does not remove the obstruction, because it reverses the Einstein term and transfers the tachyonic sign to the scalar sector.

We also investigate the theory away from Minkowski spacetime. In four dimensions, every Ricci-flat metric solves the local bulk equations through quadratic curvature order, as follows from the standard structure of quadratic gravity. The GfE expansion then fixes the cubic Weyl-operator trace as the first possible generic correction on a Ricci-flat background. A distinguished square-zero subclass behaves more strongly: Ricci-flat pp-waves are exact local vacuum solutions of the analytic metric-only logarithmic branch, and the logarithm, resolvent, and local Hessian reduce to finite expressions. This exposes the connection between GfE and the known universal higher-curvature properties of type-N spacetimes \cite{Hervik:2013cla,Hervik:2017sdr}.

The gravitational-wave branches are described through their gauge-invariant tidal response. The massless transverse-traceless sector retains the standard plus and cross polarizations, while the scalar produces the familiar mixed breathing--longitudinal pattern with a mass fixed by the GfE coupling \cite{Eardley:1973br,Eardley:1973zuo,Alves:2023rxs}. On the isolated massless branch, the averaged quadratic flux has the standard general-relativistic normalization \cite{Isaacson:1968hbi,Isaacson:1968zza}. The additional spin-2 sector is instead tachyonic and should not be interpreted as a stable massive five-polarization particle.

The paper is organized as follows. Section~\ref{sec:action} introduces the vacuum logarithmic action, the constrained $G$-field formulation, the analytic domain of the matrix logarithm, and the distinction between covariant and diagonal reductions. Section~\ref{sec:quadratic} evaluates the direct-sum curvature traces, derives their diagonal eigenvalue representation, identifies the quadratic-gravity Hessian, and establishes the Ricci-flat hierarchy. Section~\ref{sec:lineareq} derives the linearized field equations, and the auxiliary $G$-field calculation provides an independent check of their coefficients. Section~\ref{sec:spectrum} determines the pole structure, residues, source response, and stability obstruction. Section~\ref{sec:polarizations} discusses the gravitational-wave branches and their detector polarizations, while Sec.~\ref{sec:Gfield} describes their constrained $G$-field dressing. The exact Ricci-flat pp-wave sector is analyzed in Sec.~\ref{sec:ppwave}.

We use signature $(-,+,+,+)$, $\hbar=c=1$, and curvature convention
\begin{equation}
R^\rho{}_{\sigma\mu\nu}
=\partial_\mu\Gamma^\rho_{\nu\sigma}
-\partial_\nu\Gamma^\rho_{\mu\sigma}
+\Gamma^\rho_{\mu\lambda}\Gamma^\lambda_{\nu\sigma}
-\Gamma^\rho_{\nu\lambda}\Gamma^\lambda_{\mu\sigma},
\end{equation}
with $R_{\mu\nu}=R^\rho{}_{\mu\rho\nu}$. On Minkowski spacetime,
\begin{equation}
\Box\equiv\eta^{\mu\nu}\partial_\mu\partial_\nu
=-\partial_t^2+\boldsymbol\nabla^2,
\end{equation}
so a healthy massive Klein--Gordon field satisfies $(\Box-m^2)\phi=0$. For $p^\mu=(\omega,\mathbf k)$, we set $k=|\mathbf k|$ and $s=-p^2=\omega^2-k^2$.

\section{Vacuum logarithmic action, branches, and conventions}
\label{sec:action}

Before expanding the action, we specify the geometric operators and the variational branch used throughout. We follow the notation and sign conventions of the foundational GfE formulation \cite{Bianconi:2024aju}.

\subsection{Form-space geometry and the entropic action}

At each spacetime point, the geometric construction underlying GfE is defined on the finite-dimensional form space
\begin{equation}
\mathcal V_x=\Lambda^0T_x^*M\oplus\Lambda^1T_x^*M\oplus\Lambda^2T_x^*M.
\end{equation}

In four dimensions its dimensions are $1+4+6=11$. The physical topological metric is
\begin{equation}
\widetilde g=1\oplus g_{\mu\nu}\oplus g^{(2)}_{\mu\nu\rho\sigma},
\qquad
 g^{(2)}_{\mu\nu\rho\sigma}
=\frac12(g_{\mu\rho}g_{\nu\sigma}-g_{\mu\sigma}g_{\nu\rho}),
\label{eq:topmetric}
\end{equation}
and its inverse and the identity on antisymmetric pairs are
\begin{align}
[g_{(2)}]^{\mu\nu\rho\sigma}
&=\frac12(g^{\mu\rho}g^{\nu\sigma}-g^{\mu\sigma}g^{\nu\rho}),
\label{eq:inverseTopmetric}\\
I_{\mu\nu}{}^{\rho\sigma}
&\equiv g^{(2)}_{\mu\nu\alpha\beta}[g_{(2)}]^{\alpha\beta\rho\sigma}
=\frac12(\delta_\mu^\rho\delta_\nu^\sigma-\delta_\mu^\sigma\delta_\nu^\rho).
\label{eq:bivectorIdentity}
\end{align}
Thus $I_{\mu\nu}{}^{\rho\sigma}X_{\rho\sigma}=X_{\mu\nu}$ for every antisymmetric tensor $X_{\rho\sigma}$. Its trace is
\begin{equation}
I_{\mu\nu}{}^{\mu\nu}
=\frac12\left(\delta_\mu^\mu\delta_\nu^\nu-\delta_\mu^\nu\delta_\nu^\mu\right)
=\frac12(16-4)=6,
\label{eq:bivectorIdentityTrace}
\end{equation}
which confirms the six-dimensional two-form space. These factors fix the ordinary $6\times6$ matrix trace used below. The external combinatorial factor that accompanies the unrestricted wedge-basis expression in the foundational paper is absorbed here into the flattened matrix representation on six independent antisymmetric pairs. Equations~\eqref{eq:inverseTopmetric} and \eqref{eq:bivectorIdentity} implement the same normalization whenever unrestricted spacetime indices are written.
The scalar block measures zero-forms, the ordinary metric measures one-forms, and $g^{(2)}$ measures bivectors or two-forms. The associated topological curvature is then given by~\cite{Bianconi:2025awa}
\begin{equation} 
 \widetilde R_{(a_0,a_1)} = a_0 R \oplus a_1 R_{\mu\nu} \oplus R_{\mu\nu\rho\sigma}. \label{eq:generalTopCurvature} 
\end{equation} 

We assume $a_0,a_1\geq0$. The two-form block is left unweighted, so that $a_0$ and $a_1$ measure the scalar and one-form contributions relative to the Riemann endomorphism. Throughout the main text we adopt the equal-weight specialization~\cite{Bianconi:2024aju} 
\begin{equation} 
 a_0=a_1=1, \label{eq:equalCurvatureWeights} 
\end{equation} 
for which 
\begin{equation} 
 \widetilde R \equiv \widetilde R_{(1,1)} = R \oplus R_{\mu\nu} \oplus R_{\mu\nu\rho\sigma}. \label{eq:topcurv} 
\end{equation} 
This direct-sum structure produces three copies of $R$ in the linear curvature trace and includes the Kretschmann invariant in the quadratic trace. The numerical coefficients and the fixed mass relation derived in the main text refer to Eq.~\eqref{eq:equalCurvatureWeights}. The corresponding formulas for general $a_0$ and $a_1$ are collected in Appendix~\ref{app:generalCurvatureWeights}.

The full GfE proposal describes bosonic matter by a Dirac--K\"ahler-like field
\begin{equation}
|\Phi\rangle
=\phi\oplus\omega_\mu\,\mathrm dx^\mu
\oplus\zeta_{\mu\nu}\,\mathrm dx^\mu\wedge\mathrm dx^\nu,
\label{eq:topologicalMatterField}
\end{equation}
where $\zeta_{\mu\nu}=-\zeta_{\nu\mu}$. With the Hodge--Dirac operator $\mathfrak D=d+\delta$, where $d$ is the exterior derivative and $\delta$ is the metric codifferential, the matter-induced operator is postulated to have the schematic form \cite{Bianconi:2024aju}
\begin{equation}
\widetilde M
=\mathfrak D|\Phi\rangle\langle\Phi|\mathfrak D
+(m^2+\xi R)|\Phi\rangle\langle\Phi|.
\label{eq:fullMatterOperator}
\end{equation}
The induced topological metric is then
\begin{equation}
\widetilde{\mathbf G}
=\widetilde g+\alpha\widetilde M-\beta\widetilde R.
\label{eq:fullInducedMetric}
\end{equation}
The foundational proposal assumes $\alpha>0$ and $\beta>0$. In four dimensions $[\alpha]=L^4$ for its normalization, while $[\beta]=L^2$, so the logarithm acts on a dimensionless mixed operator. Whenever $\beta<0$ is mentioned below, it is only a formal analytic continuation of the metric-only action outside the parameter domain postulated in the original construction.

The local action is
\begin{equation}
S[g,\Phi]
=-\frac{1}{\mpl^4}\int\mathrm d^4x\sqrt{-g}\,
\Tr_F\operatorname{Log}\!\left(\widetilde{\mathbf G}\widetilde g^{-1}\right),
\label{eq:fullGfEAction}
\end{equation}
where $\mpl$ is the ordinary Planck length, so that $\mpl^2=G_N$ in units $\hbar=c=1$. At weak coupling the first logarithmic term gives Einstein--Hilbert gravity plus the standard quadratic matter action. For a purely scalar topological field the leading Lagrangian is
\begin{equation}
\mathcal L_{\rm weak}
=\frac{1}{\mpl^4}\left[
3\beta R-\alpha\nabla_\mu\bar\phi\nabla^\mu\phi
-\alpha(m^2+\xi R)|\phi|^2+\cdots\right].
\label{eq:weakMatterLimit}
\end{equation}
This limit explains why $\beta$ fixes the vacuum gravitational normalization. It also shows that the matter sector is not merely an external source: the operator $\widetilde M$ in Eq.~\eqref{eq:fullMatterOperator} depends on the metric through the curvature scalar, the Hodge--Dirac operator, and the form-space pairings. Consequently, a background with $\bar\Phi\neq0$ can modify the effective gravitational coefficients and introduce linear metric--matter mixing. In particular, the nonminimal term $\xi R|\Phi\rangle\langle\Phi|$ can alter the metric Hessian on such a background. Whether this modification removes, shifts, or constrains any of the vacuum poles cannot be inferred from the weak-coupling expression alone and requires the complete coupled perturbation analysis.

Throughout, $\Tr_0$, $\Tr_1$, and $\Tr_2$ denote the ordinary matrix traces on $\Lambda^0$, $\Lambda^1$, and $\Lambda^2$, respectively, while $\Tr_F\equiv\Tr_0+\Tr_1+\Tr_2$ denotes the trace on their direct sum. The notation $\Tr_{\Lambda^2}$ is synonymous with $\Tr_2$ and is used when emphasizing the two-form curvature endomorphism. An unsubscripted $\Tr$ denotes a generic finite-dimensional matrix trace. Moreover, by the \emph{metric-only vacuum branch} we mean the sector obtained by setting the topological matter field to zero and eliminating the auxiliary $G$-field through its algebraic equation, so that $g_{\mu\nu}$ is the only independent field. The logarithmic action and its auxiliary-$G$ rewriting represent the same dynamics on this branch only when the algebraic equation and the same boundary conditions are imposed.

Therefore the metric--matter Hessian about
$(\bar g_{\mu\nu},\bar\Phi)=(\eta_{\mu\nu},0)$ contains no term
linear in both $h_{\mu\nu}$ and $\delta\Phi$. The vacuum
gravitational Hessian derived below is thus also the gravitational
block of the complete linearized theory about this background. Matter
can alter that block through Eq.~\eqref{eq:fullInducedMetric} only
when the background matter configuration is nonzero, or when a
different constraint or phase-space prescription is adopted.

\subsection{Metric-only vacuum branch}

The vacuum sector is obtained by setting $|\Phi\rangle=0$, so $\widetilde M=0$ and
\begin{equation}
\widetilde{\mathbf G}=\widetilde g-\beta\widetilde R.
\label{eq:inducedmetric}
\end{equation}

Equation~\eqref{eq:inducedmetric} is obtained only after imposing $\Phi=0$. Before this restriction, $\widetilde M$ depends on the metric through the Hodge--Dirac operator, the form-space pairings, and the nonminimal curvature coupling in Eq.~\eqref{eq:fullMatterOperator}. Nevertheless, because
$\widetilde M$ is bilinear in $\Phi$, its variations at
$(\bar g_{\mu\nu},\bar\Phi)=(\eta_{\mu\nu},0)$ satisfy
\begin{equation}
\left.\delta_g\widetilde M\right|_{\bar\Phi=0} = \left.\delta_\Phi\widetilde M\right|_{\bar\Phi=0} =0,
\end{equation}
and
\begin{equation}
\left.\delta_g^2\widetilde M\right|_{\bar\Phi=0} = \left.\delta_g\delta_\Phi\widetilde M\right|_{\bar\Phi=0} =0,
\qquad
\left.\delta_\Phi^2\widetilde M\right|_{\bar\Phi=0} \neq0
\end{equation}
in general. Thus the pure metric block and the mixed metric--matter block of the reduced Hessian are unaffected by $\widetilde M$, whereas the matter block remains nontrivial. After imposing Eq.~\eqref{eq:vacuumGfieldEquation} and eliminating the constrained $G$-field, the vacuum metric Hessian derived below is therefore the gravitational block of the reduced coupled theory about the zero-matter Minkowski background.

The metric-only vacuum action is given by
\begin{equation}
S_{\rm GfE}[g]
=-\frac{1}{\mpl^4}\int\mathrm d^4x\sqrt{-g}\,
\Tr_F\operatorname{Log}\!\left(\widetilde{\mathbf G}\widetilde g^{-1}\right).
\label{eq:originalaction}
\end{equation}

We now define the direct-sum curvature endomorphism
\begin{equation}
\mathbb R\equiv\widetilde R\widetilde g^{-1} =R\oplus R_\mu{}^\nu\oplus\mathcal R_{\mu\nu}{}^{\rho\sigma},
\label{eq:Roperator}
\end{equation}
where
\begin{equation}
\mathcal R_{\mu\nu}{}^{\rho\sigma} =R_{\mu\nu\alpha\beta}[g_{(2)}]^{\alpha\beta\rho\sigma} =R_{\mu\nu}{}^{\rho\sigma}.
\label{eq:bivectorop}
\end{equation}

The two-form sector is six-dimensional in four spacetime dimensions.
To make the trace convention explicit, we introduce a collective
antisymmetric-pair index
\begin{equation}
A=[\mu\nu],
\qquad
\mu<\nu,
\end{equation}
so that an endomorphism of $\Lambda^2T_x^*M$ may be represented as
an ordinary $6\times6$ matrix $\mathcal X_A{}^B$. Its trace is
therefore
\begin{equation}
\Tr_2\mathcal X
=
\sum_{A=1}^{6}\mathcal X_A{}^A.
\label{eq:twoFormTraceMatrix}
\end{equation}
When the same trace is written using unrestricted spacetime indices,
it takes the form
\begin{equation}
\Tr_2\mathcal X
\equiv
\mathcal X_{\mu\nu}{}^{\mu\nu}.
\label{eq:twoFormTraceConvention}
\end{equation}
No additional factor of $1/2$ is required in
Eq.~\eqref{eq:twoFormTraceConvention}. The antisymmetric-pair
normalization and the removal of the double counting between
$(\mu,\nu)$ and $(\nu,\mu)$ are already encoded in the factors of
$1/2$ appearing in the induced two-form metric, its inverse, and the
identity operator in Eqs.~\eqref{eq:topmetric},
\eqref{eq:inverseTopmetric}, and \eqref{eq:bivectorIdentity}.
Consequently, Eq.~\eqref{eq:twoFormTraceConvention} is precisely the
ordinary trace of the corresponding $6\times6$ matrix on the space
of independent two-forms.

We denote the identity endomorphism on the full direct-sum space
$\mathcal V_x=\Lambda^0\oplus\Lambda^1\oplus\Lambda^2$ by
$\widetilde I$. Since
\begin{equation}
\widetilde{\mathbf G}\widetilde g^{-1}
=
\widetilde I-\beta\mathbb R
\end{equation}
on the metric-only vacuum branch, the action can be written compactly
as
\begin{equation}
S_{\rm GfE}[g]
=
-\frac{1}{\mpl^4}
\int\mathrm d^4x\sqrt{-g}\,
\Tr_F\operatorname{Log}
\!\left(\widetilde I-\beta\mathbb R\right).
\label{eq:compactaction}
\end{equation}

\subsection{Covariant operator formulation and diagonal reductions}
\label{sec:diagonalConventions}

Equation~\eqref{eq:compactaction} is basis independent and does not assume that either $R_\mu{}^\nu$ or $\mathcal R_{\mu\nu}{}^{\rho\sigma}$ is diagonalizable. This is the general operator formulation of the metric-only branch~\cite{Bianconi:2024aju,higham2008functions}. It is useful to distinguish it from two related but logically different specializations.

First, we assume that the one-form and two-form curvature endomorphisms are diagonalizable over $\mathbb C$ at a given spacetime point. We denote their eigenvalues, counted with algebraic multiplicity, by
\begin{equation}
\{r_i\}_{i=1}^{4},
\qquad
\{\lambda_A\}_{A=1}^{6}.
\label{eq:curvatureEigenvalues}
\end{equation} In
bases adapted separately to the one-form and two-form sectors,
\begin{equation}
R_\mu{}^\nu\sim\operatorname{diag}(r_1,\ldots,r_4),
\qquad
\mathcal R_A{}^B\sim
\operatorname{diag}(\lambda_1,\ldots,\lambda_6).
\label{eq:diagonalizableCurvatureBlocks}
\end{equation}
On compatible scalar branches, the trace-log is then
\begin{align}
\Tr_F\operatorname{Log}
(\widetilde I-\beta\mathbb R)
={}&
\log(1-\beta R)
+\sum_{i=1}^{4}\log(1-\beta r_i)
\notag\\
&+\sum_{A=1}^{6}\log(1-\beta\lambda_A).
\label{eq:diagonalEigenvalueLog}
\end{align}
This is not a new action: it is the eigenvalue representation of Eq.~\eqref{eq:compactaction}. The principal-logarithm condition \eqref{eq:principal-domain} reduces in this representation to the requirement that none of
\begin{equation}
1-\beta R,
\qquad
1-\beta r_i,
\qquad
1-\beta\lambda_A
\end{equation}
lies on the closed negative real axis. Because the curvature endomorphisms are real, their nonreal eigenvalues occur in complex-conjugate pairs. If the principal-logarithm condition is satisfied, the principal logarithm of the corresponding real matrix is itself real~\cite{higham2008functions}.

Second, some applications restrict the metric to a symmetry class for which the flattened curvature matrices are diagonal in a fixed coordinate-adapted basis~\cite{Thattarampilly:2025krv,Thattarampilly:2026wsw}. We can legitimately substitute such a background into the field equations obtained from the full covariant action. By contrast, if we substitute the diagonal ansatz into the action \emph{before} variation, we define a reduced variational problem whose equivalence to the restricted covariant equations is not automatic \cite{Palais:1979rca,Fels:2001rv}. We denote the retained variables by $q_\parallel$ and the discarded off-diagonal variables by $q_\perp$. This reduction is equivalent to the full theory only if
\begin{equation}
\left. \frac{\delta S}{\delta q_\perp}\right|_{q_\perp=0} =0
\label{eq:consistentDiagonalTruncation}
\end{equation}
identically on the reduced configuration space. We do not assume Eq.~\eqref{eq:consistentDiagonalTruncation} in the perturbative analysis below. Instead, we vary the full covariant operator and impose background symmetries only afterwards.

The distinction is immaterial for invariant trace values on a diagonalizable background, but it is essential for perturbations and for non-diagonalizable geometries. In particular, a nonzero square-zero curvature endomorphism cannot be diagonalized. Indeed, if $\mathcal R^2=0$, then every eigenvalue of $\mathcal R$ vanishes; if $\mathcal R$ were diagonalizable, it would therefore be similar to the zero matrix and hence would itself vanish \cite{Horn_Johnson_2012}.

\subsection{The auxiliary \texorpdfstring{$G$}{G}-field and the metric-only branch}

The logarithm can be rewritten by introducing a topological $G$-field $\widetilde{\mathcal G}$ and a constraint field. Its algebraic equation is
\begin{equation}
\widetilde{\mathcal G}^{-1} =\widetilde I+\alpha\widetilde M\widetilde g^{-1} -\beta\widetilde R\widetilde g^{-1},
\label{eq:generalGfieldEquation}
\end{equation}
and in vacuum
\begin{equation}
\widetilde{\mathcal G}^{-1}=\widetilde I-\beta\mathbb R.
\label{eq:vacuumGfieldEquation}
\end{equation}

The gravitational part of the enlarged action has a dressed Einstein--Hilbert form with
\begin{equation}
R_{\mathcal G}=\Tr_F(\widetilde g^{-1}\widetilde{\mathcal G}\widetilde R),
\qquad
\Lambda_{\mathcal G} =\frac{1}{2\beta}\Tr_F(\widetilde{\mathcal G}-\widetilde I-\operatorname{Log}\widetilde{\mathcal G}).
\label{eq:dressedScalars}
\end{equation}
In the foundational convention, the vacuum metric equation is written as~\cite{Bianconi:2024aju}
\begin{equation}
R^{\mathcal G}_{(\mu\nu)} -\frac12g_{\mu\nu}(R_{\mathcal G}-2\Lambda_{\mathcal G}) +D_{(\mu\nu)}=0,
\label{eq:dressedVacuumEquation}
\end{equation}
where $D_{\mu\nu}$ contains second derivatives of the scalar, one-form, and two-form blocks of $\widetilde{\mathcal G}$ \cite{Bianconi:2024aju}. Equations~\eqref{eq:dressedScalars} and \eqref{eq:dressedVacuumEquation} fix the notation used throughout the auxiliary-field check below.

There are also two logically distinct uses of the auxiliary rewriting. If Eq.~\eqref{eq:vacuumGfieldEquation} is enforced and $\widetilde{\mathcal G}$ is eliminated, the result is exactly the metric-only logarithmic action and the metric equation is generically fourth order. If $\widetilde{\mathcal G}$ is instead assigned an independent phase space, the enlarged equations are second order but require a separate Hamiltonian and constraint analysis to determine their physical modes. The spectrum derived below belongs to the first, metric-only interpretation. We later verify that linearizing Eq.~\eqref{eq:dressedVacuumEquation} in these conventions and then eliminating the $G$-field reproduces the same local bulk Minkowski Hessian.

\subsection{Analytic matrix logarithm and Lorentzian domain}
\label{sec:logdomain}

The pairing induced by $g_{\mu\nu}$ on one-forms is Lorentzian, while the induced pairing on two-forms in four spacetime dimensions has split signature $(3,3)$. Consequently, the direct-sum form space is an indefinite inner-product space rather than a positive-definite Hilbert space. Although the mixed Ricci and Riemann endomorphisms are self-adjoint with respect to the corresponding indefinite pairings, indefinite self-adjointness does not imply diagonalizability or a real spectrum, and the ordinary Hermitian spectral theorem does not apply \cite{bognar1974indefinite}.

It is therefore necessary to distinguish the following structures:
\begin{enumerate}
\item the entropy interpretation in terms of positive operators on a
positive-definite Hilbert space;
\item the Lorentzian index contractions defining the mixed geometric
endomorphism;
\item the analytic matrix function
$\Tr_F\operatorname{Log}(\widetilde I-\beta\mathbb R)$.
\end{enumerate}
The Minkowski perturbative calculation requires only the analytic germ of the third structure at the identity. The resulting analytic continuation does not, by itself, establish that all admitted configurations belong to the positive-operator domain required by the entropy interpretation.

We define
\begin{equation}
\mathbb A\equiv\widetilde I-\beta\mathbb R.
\label{eq:Adefinition}
\end{equation}
Throughout this work, $\operatorname{Log}\mathbb A$ denotes the principal matrix logarithm, namely the primary logarithm induced by the scalar branch analytic on
$\mathbb C\setminus(-\infty,0]$ and normalized by
\begin{equation}
\operatorname{Log}\widetilde I=0.
\label{eq:logIdentityNormalization}
\end{equation}

The principal logarithm exists if and only if
\begin{equation}
\sigma(\mathbb A)\cap(-\infty,0]=\varnothing.
\label{eq:principal-domain}
\end{equation}
Moreover, it is the unique matrix $X$ satisfying
\begin{equation}
e^X=\mathbb A,
\qquad
-\pi<\operatorname{Im}\lambda<\pi
\quad
\text{for every }\lambda\in\sigma(X).
\end{equation}
No diagonalizability assumption is required. Moreover, if $\mathbb A$ is real and satisfies Eq.~\eqref{eq:principal-domain}, then $\operatorname{Log}\mathbb A$ is real \cite{higham2008functions}.

For any matrix logarithm of an invertible matrix, we have
\begin{equation}
\det\mathbb A=\exp\!\left[\Tr_F\operatorname{Log}\mathbb A\right],
\label{eq:traceLogDetIdentity}
\end{equation}
and consequently, for any selected scalar logarithm we have
\begin{equation}
\Tr_F\operatorname{Log}\mathbb A-\log(\det\mathbb A)\in 2\pi\mathrm{i}\mathbb Z.
\label{eq:traceLogDetModulo}
\end{equation}

On a sufficiently small connected neighborhood of $\widetilde I$, with the matrix and scalar logarithms chosen compatibly and normalized by
$\operatorname{Log}\widetilde I=0$ and $\log 1=0$, this becomes
\begin{equation}
\Tr_F\operatorname{Log}\mathbb A = \log\det\mathbb A.
\label{eq:traceLogDetLocalBranch}
\end{equation}
The qualification concerning compatible branches is essential: outside such a local domain, the trace of the principal matrix logarithm need not equal the principal scalar logarithm of the determinant.

More generally, if $f$ is analytic on a neighborhood of the spectrum $\sigma(\mathbb A)$, then
\begin{equation}
\Tr_F f(\mathbb A)=\sum_i f(\lambda_i),
\label{eq:traceFunctionSpectralValue}
\end{equation}
where the eigenvalues $\lambda_i$ are counted with algebraic multiplicity. Thus the trace of an analytic matrix function depends only on the eigenvalues and their algebraic multiplicities, and not on the nilpotent parts of the Jordan blocks. The full matrix $f(\mathbb A)$, however, generally retains information about the Jordan structure.

For variations that remain within the chosen analytic domain,
Jacobi's formula gives
\begin{equation}
\delta\Tr_F\operatorname{Log}\mathbb A=\Tr_F\!\left(\mathbb A^{-1}\delta\mathbb A\right).
\label{eq:traceLogFirstVariation}
\end{equation}
This identity does not require
$[\mathbb A,\delta\mathbb A]=0$. It should not be confused with the matrix identity $\delta\operatorname{Log}\mathbb A=\mathbb A^{-1}\delta\mathbb A$, which is generally false for noncommuting variations. The variational problem therefore retains the full resolvent
\begin{equation}
\widetilde{\mathcal G} = \mathbb A^{-1},
\end{equation}
even though the numerical trace depends only on the eigenvalues. The pp-wave sector in Sec.~\ref{sec:ppwave} makes this distinction explicit.

A sufficient condition for convergence of the Taylor expansion about
the identity is
\begin{equation}
\rho(\beta\mathbb R)<1,
\label{eq:spectral-radius}
\end{equation}
where
\begin{equation}
\rho(\beta\mathbb R) = \max_{\lambda\in\sigma(\beta\mathbb R)}|\lambda|
\end{equation}
is the spectral radius. The stronger condition
\begin{equation}
\|\beta\mathbb R\|<1
\end{equation}
in any specified submultiplicative matrix norm is also sufficient. The Taylor domain is a strict subset of the principal-logarithm domain:
\begin{equation}
\left\{
X:\rho(X)<1
\right\}
\subsetneq
\left\{
X:
\sigma(\widetilde I-X)\cap(-\infty,0]=\varnothing
\right\}.
\label{eq:TaylorPrincipalInclusion}
\end{equation}

At Minkowski spacetime,
\begin{equation}
\mathbb R=0,
\qquad
\mathbb A=\widetilde I,
\qquad
\operatorname{Log}\mathbb A=0
\end{equation}
on the principal branch fixed by
Eq.~\eqref{eq:logIdentityNormalization}. The principal-logarithm
condition is a spectral branch condition and should not be identified
with positivity in the quantum-information sense: a matrix may
satisfy Eq.~\eqref{eq:principal-domain} without being Hermitian or
positive definite.

Accordingly, all perturbative and exact results below are statements
about the analytic metric-only continuation defined by the principal
matrix logarithm. For a square-zero curvature operator,
\begin{equation}
\mathbb R^2=0,
\end{equation}
we have
\begin{equation}
\sigma(\widetilde I-\beta\mathbb R)=\{1\},
\qquad
\operatorname{Log}
(\widetilde I-\beta\mathbb R)
=
-\beta\mathbb R,
\end{equation}
independently of diagonalizability and without requiring
$\|\beta\mathbb R\|<1$. This is the analytic functional calculus
used in the exact pp-wave analysis of
Sec.~\ref{sec:ppwave}.

\section{Covariant vacuum reduction through quadratic curvature order}
\label{sec:quadratic}

With the analytic branch and operator normalization fixed, we now determine the local curvature content selected by the first two powers of the logarithm. The calculation is fully covariant and precedes any weak-field decomposition. Curvature order is kept distinct from perturbative order: the expressions through $\mathbb R^2$ remain nonlinear in the metric, and only after the reduction is complete are they specialized to the Minkowski Hessian.

\subsection{Bivector normalization and constant-curvature check}
\label{sec:constantCurvatureCheck}

A useful convention check is a four-dimensional constant-curvature geometry,
\begin{equation}
R_{\mu\nu\rho\sigma}
=K(g_{\mu\rho}g_{\nu\sigma}-g_{\mu\sigma}g_{\nu\rho})
=2K g^{(2)}_{\mu\nu\rho\sigma}.
\label{eq:constantCurvatureRiemann}
\end{equation}
Equations~\eqref{eq:inverseTopmetric} and \eqref{eq:bivectorIdentity} then give
\begin{equation}
\mathcal R_{\mu\nu}{}^{\rho\sigma}=2K I_{\mu\nu}{}^{\rho\sigma}.
\label{eq:constantCurvatureOperator}
\end{equation}
Hence the two-form curvature endomorphism has eigenvalue $2K$ with multiplicity six. Its ordinary matrix traces are
\begin{align}
\Tr_2\mathcal R&=6(2K)=12K=R,\\
\Tr_2\mathcal R^2&=6(2K)^2=24K^2=R_{\mu\nu\rho\sigma}R^{\mu\nu\rho\sigma}.
\label{eq:constantCurvatureTraceCheck}
\end{align}
This check fixes the factors of two that otherwise vary among bivector conventions.

\subsection{First topological trace}

The direct-sum trace is the sum of the traces in the zero-, one-, and two-form sectors. The scalar block gives $R$. The one-form block gives
\begin{equation}
\Tr_{1}(R_{\mu}{}^{\nu})=R_{\mu}{}^{\mu}=R.
\end{equation}
For the two-form block,
\begin{align}
\Tr_{2}(\cR)
&=R_{\mu\nu\rho\sigma}[g_{(2)}]^{\rho\sigma\mu\nu}\nonumber\\
&=\frac12R_{\mu\nu\rho\sigma}
\left(g^{\rho\mu}g^{\sigma\nu}-g^{\rho\nu}g^{\sigma\mu}\right)
=R.
\end{align}
The first contraction equals $R$, whereas the second equals $-R$ by antisymmetry in the last Riemann index pair. The explicit minus sign between them therefore gives $\tfrac12[R-(-R)]=R$. Hence
\begin{equation}
\TrF\mathbb R=3R.
\label{eq:trace1}
\end{equation}
This is the origin of the factor $3$ in the Einstein limit of GfE. A detailed derivation of the two-form trace normalization, including the ordered-pair convention and the associated factors of one half, is given in Appendix~\ref{app:trace}.

\subsection{Second topological trace}

The scalar and vector sectors immediately give
\begin{equation}
\Tr_{0}(\mathbb R^{2})=R^{2},
\qquad
\Tr_{1}(\mathbb R^{2})=R_{\mu\nu}R^{\mu\nu}.
\end{equation}
For the two-form sector,
\begin{align}
\Tr_{2}(\cR^{2})
&=R_{\mu\nu\alpha\beta}
[g_{(2)}]^{\alpha\beta\rho\sigma}
R_{\rho\sigma\gamma\delta}
[g_{(2)}]^{\gamma\delta\mu\nu}
\nonumber\\
&=R_{\mu\nu}{}^{\rho\sigma}R_{\rho\sigma}{}^{\mu\nu}
=R_{\mu\nu\rho\sigma}R^{\mu\nu\rho\sigma}.
\end{align}
Hence
\begin{equation}
\TrF\mathbb R^{2}
=R^{2}+R_{\mu\nu}R^{\mu\nu}
+R_{\mu\nu\rho\sigma}R^{\mu\nu\rho\sigma}.
\label{eq:trace2}
\end{equation}
The two-form contraction in this result, and in particular the absence of an additional factor of $1/2$, is shown explicitly in Appendix~\ref{app:trace}.

\subsection{Diagonalizable subcase and invariant recovery}
\label{sec:diagonalTraceSubcase}

The diagonalizable representation in Eq.~\eqref{eq:diagonalEigenvalueLog} provides a direct check of the covariant traces. We expand each scalar logarithm about the identity and obtain
\begin{widetext}
\begin{align}
-\Tr_F\operatorname{Log}
(\widetilde I-\beta\mathbb R)={}&\beta\left(R+\sum_i r_i+\sum_A\lambda_A\right)+\frac{\beta^2}{2}\left(R^2+\sum_i r_i^2+\sum_A\lambda_A^2
\right)+\order(\beta^3\mathbb R^3).
\label{eq:diagonalLogExpansion}
\end{align}
The spectral sums are invariant traces,
\begin{align}
\sum_i r_i&=\Tr_1 R_\mu{}^\nu=R,
\qquad
\sum_A\lambda_A=\Tr_2\mathcal R=R,
\label{eq:diagonalFirstSpectralSums}\\
\sum_i r_i^2&=\Tr_1(R_\mu{}^\nu)^2=R_{\mu\nu}R^{\mu\nu},
\qquad
\sum_A\lambda_A^2=\Tr_2\mathcal R^2=R_{\mu\nu\rho\sigma}R^{\mu\nu\rho\sigma}.
\label{eq:diagonalSecondSpectralSums}
\end{align}
\end{widetext}
Consequently, the diagonalizable subcase reproduces Eqs.~\eqref{eq:trace1} and \eqref{eq:trace2} exactly. It therefore leads to the same curvature-truncated action, the same Minkowski Hessian, and the same nonzero pole locations derived below. No coefficient changes merely because the curvature operator is evaluated in an eigenbasis.

A strict diagonal restriction on the perturbations is different. In a fixed independent-pair basis,
\begin{equation}
\Tr_2[(\delta\mathcal R)^2]=\sum_A(\delta\mathcal R_A{}^A)^2+\sum_{A\neq B}\delta\mathcal R_A{}^B\delta\mathcal R_B{}^A.
\label{eq:diagonalOffDiagonalSplit}
\end{equation}

Requiring $\delta\mathcal R_A{}^B=0$ for $A\neq B$ removes the second sum by restricting the allowed metric perturbations. The remaining quadratic form is the pullback of the covariant Hessian to that restricted subspace, but it is not the complete Hessian and cannot be used to count all spin-projector degrees of freedom.

The same distinction is visible on a diagonalizable Ricci-flat background. Since the scalar and one-form blocks vanish and
$\sum_A\lambda_A=0$, the exact local density becomes
\begin{equation}
	-\Tr_F\operatorname{Log}(\widetilde I-\beta\mathbb R)=-\sum_{A=1}^{6}\log(1-\beta\lambda_A),
\label{eq:diagonalRicciFlatLog}
\end{equation}
with expansion
\begin{equation}
\frac{\beta^2}{2}\sum_A\lambda_A^2+\frac{\beta^3}{3}\sum_A\lambda_A^3+\order(\beta^4\mathcal C^4).
\label{eq:diagonalRicciFlatExpansion}
\end{equation}
The quadratic density need not vanish; on a Ricci-flat metric it is
the Kretschmann invariant. Nevertheless, after varying the full
four-dimensional covariant action, its complete quadratic
Euler--Lagrange tensor vanishes by the Gauss--Bonnet reduction in
Sec.~\ref{sec:ricciFlatProtection}. Thus a nonzero diagonal
trace-log density must not be confused with a nonzero correction to
the local bulk equations. The first possible generic covariant
correction to a Ricci-flat background remains the cubic spectral
sum $\sum_A\lambda_A^3=\Tr_{\Lambda^2}(\mathcal C^3)$.

\subsection{Curvature truncation and the exact local bulk Minkowski Hessian}

We write
\begin{equation}
g_{\mu\nu}=\eta_{\mu\nu}+h_{\mu\nu}.
\end{equation}
Every curvature tensor has an expansion
\begin{equation}
R_{\mu\nu\rho\sigma}=R^{(1)}_{\mu\nu\rho\sigma}[h]+R^{(2)}_{\mu\nu\rho\sigma}[h,h]+\cdots,
\end{equation}
so $\mathbb R=\order(h)$. It follows that
\begin{equation}
\TrF(\mathbb R^{n})=\order(h^{n}).
\end{equation}
The logarithmic and perturbative expansions should be distinguished. We now define the complete nonlinear action through curvature-squared order by
\begin{align}
S_{\leq2,\mathrm{curv}}={}&\frac{1}{\mpl^{4}}\int \dd^{4}x\sqrt{-g}\,\bigg\{3\beta R+\frac{\beta^{2}}{2}\bigg[R^{2}+R_{\mu\nu}R^{\mu\nu}
\notag\\
&\hspace{2.8cm}
+R_{\mu\nu\rho\sigma}R^{\mu\nu\rho\sigma}
\bigg]\bigg\}.
\label{eq:curvquad}
\end{align}
The full analytic action is
\begin{equation}
S_{\rm GfE}=S_{\leq2,\mathrm{curv}}+\order(\mathbb R^3),
\label{eq:curvatureOrderExpansion}
\end{equation}
where the displayed first term remains nonlinear in the metric and contains arbitrarily high powers of $h_{\mu\nu}$ when expanded away from a background. Separately, around Minkowski we have $\mathbb R=\order(h)$, so
\begin{equation}
S_{\rm GfE}^{(2)}[h]
=\left(S_{\leq2,\mathrm{curv}}\right)^{(2)}[h],
\qquad
\order(\mathbb R^3)=\order(h^3).
\label{eq:MinkowskiHessianIdentity}
\end{equation}

Thus the first two curvature orders determine the complete local bulk Minkowski Hessian of the assumed continuum action. This algebraic statement is distinct from the derivative-domain question of whether modes near the nonzero poles belong to a low-energy EFT.

In four dimensions, the Euler density is \cite{Lovelock:1971yv}
\begin{equation}
E_{4}=R_{\mu\nu\rho\sigma}R^{\mu\nu\rho\sigma}
-4R_{\mu\nu}R^{\mu\nu}+R^{2},
\label{eq:GB}
\end{equation}
and therefore one has
\begin{equation}
R^{2}+R_{\mu\nu}R^{\mu\nu}+R_{\mu\nu\rho\sigma}R^{\mu\nu\rho\sigma}
=E_{4}+5R_{\mu\nu}R^{\mu\nu}.
\label{eq:GBreduce}
\end{equation}

After discarding the Euler density, which does not contribute to the
local four-dimensional bulk equations, the complete nonlinear
curvature-truncated action is locally bulk-equivalent to
\begin{equation}
S_{\leq2,\mathrm{curv}}
\doteq
\frac{1}{\mpl^{4}}
\int \dd^{4}x\sqrt{-g}
\left[3\beta R+\frac{5\beta^{2}}{2}R_{\mu\nu}R^{\mu\nu}\right],
\label{eq:reducedaction}
\end{equation}
where $\doteq$ denotes equality of local four-dimensional bulk equations under boundary conditions for which the Euler variation does not contribute. We take the second perturbative variation of Eq.~\eqref{eq:reducedaction} about Minkowski and obtain the exact Hessian in Eq.~\eqref{eq:MinkowskiHessianIdentity}. It is at this point, and not before the direct-sum traces have been evaluated, that the flat-space problem is identified as a specific one-parameter member of four-derivative quadratic gravity. The general flat-space spectrum, the second-order auxiliary-field description, and the residue structure are classical results \cite{Stelle:1976gc,Stelle:1977ry,Whitt:1984pd,Hindawi:1995an,Salvio:2018crh}. The explicit $R^{2}$ coefficient cancels in the Ricci-squared basis, but $R_{\mu\nu}R^{\mu\nu}$ still carries a scalar trace mode.

This completes the GfE-specific determination of the Minkowski Hessian. The remaining flat-space mode analysis is that of standard four-dimensional quadratic gravity with the coefficients $A$ and $B$ fixed as above. We therefore use the established spin decomposition and pole formulas, retaining only the intermediate steps needed to verify our signature, normalization, and source conventions.

\subsection{Location in quadratic-gravity parameter space}
\label{sec:qgmap}

To display what is standard and what is fixed by GfE, we use the four-dimensional identity
\begin{equation}
R_{\mu\nu}R^{\mu\nu}
=\frac12 C_{\mu\nu\rho\sigma}C^{\mu\nu\rho\sigma}
+\frac13R^2-\frac12E_4,
\label{eq:RicciWeylMap}
\end{equation}
where $C_{\mu\nu\rho\sigma}$ is the Weyl tensor. After dropping the four-dimensional Euler-density term, because it does not change the local bulk field equations, Eq.~\eqref{eq:reducedaction} is therefore
\begin{align}
S_{\rm quad}={}&\int \dd^4x\sqrt{-g}\bigg[\frac{M_{\rm eff}^2}{2}R+a_{\rm GfE}R^2\notag\\
&\hspace{2.3cm}+b_{\rm GfE}C_{\mu\nu\rho\sigma}C^{\mu\nu\rho\sigma}
\bigg].
\label{eq:GfEQuadraticPlane}
\end{align}
with
\begin{equation}
 \frac{M_{\rm eff}^2}{2} \equiv \frac{3\beta}{\mpl^4},
 \quad
a_{\rm GfE}=\frac{5\beta^2}{6\mpl^4},
\quad
b_{\rm GfE}=\frac{5\beta^2}{4\mpl^4},
\quad
\frac{b_{\rm GfE}}{a_{\rm GfE}}=\frac32.
\label{eq:GfECouplingRatio}
\end{equation}

Generic quadratic gravity allows $a$ and $b$ to vary independently. GfE instead selects the one-dimensional ray $b/a=3/2$ and ties both coefficients to the same parameter that fixes the Einstein term. The ensuing projector algebra is therefore standard, whereas the mass relation and the simultaneous sign constraints follow from the GfE coefficient locking.

For direct comparison with a common standard convention, we can write
\begin{equation}
S_{\rm std}=\int\!\dd^4x\sqrt{-g}\left[\frac{M^2}{2}R+aR^2+bC_{\mu\nu\rho\sigma}C^{\mu\nu\rho\sigma}\right],
\label{eq:standardQuadraticConvention}
\end{equation}
with the same curvature and signature conventions as above. Table~\ref{tab:dictionary} supplies the complete dictionary used in this paper.
\begin{table*}[t]
\centering
\caption{Convention dictionary between the GfE coefficients and the standard quadratic-gravity basis in Eq.~\eqref{eq:standardQuadraticConvention}, modulo the Euler density. The mass formulas follow from the established flat-space quadratic-gravity spectrum in this convention.}
\label{tab:dictionary}
\begin{ruledtabular}
\begin{tabular}{lll}
Quantity & GfE / Ricci-squared basis & Standard $R^2$--$C^2$ basis\\
\hline
Einstein coefficient & $A=3\beta/\mpl^4$ & $M^2=2A$\\
Quadratic coefficient & $B=5\beta^2/(2\mpl^4)$ in $BR_{\mu\nu}^2$ & $a=B/3$, $b=B/2$\\
Scalar pole & $m_0^2=A/(2B)$ & $m_0^2=M^2/(12a)$\\
Spin-2 pole & $m_2^2=-A/B$ & $m_2^2=-M^2/(4b)$\\
Newton coupling & $G_{\rm eff}=1/(16\pi A)$ & $G_{\rm eff}=1/(8\pi M^2)$\\
\end{tabular}
\end{ruledtabular}
\end{table*}

We define
\begin{equation}
A\equiv\frac{3\beta}{\mpl^{4}},
\qquad
B\equiv\frac{5\beta^{2}}{2\mpl^{4}},
\label{eq:AB}
\end{equation}
so that the bulk quadratic-curvature action is $\int\sqrt{-g}(AR+BR_{\mu\nu}^{2})$.
The effective Newton coupling determined by the Einstein term is 
\begin{equation}
 G_{\rm eff} = \frac{1}{16\pi A} = \frac{1}{8\pi M_{\rm eff}^2} = \frac{\mpl^4}{48\pi\beta}. \label{eq:Geff}
\end{equation}
With the conventional positive sign of the matter action, a positive-energy massless graviton and attractive positive Newton coupling require
\begin{equation}
\beta>0.
\label{eq:betapositive}
\end{equation}
The pure-gravity overall sign would be conventional in isolation, but it is physical once the relative sign to positive-energy matter is fixed. Matching the low-energy gravitational coupling to the measured Newton constant, 
\begin{equation} 
 G_{\rm eff}=G_N, \end{equation} then fixes \begin{equation} \beta = \frac{\mpl^2}{48\pi}. \label{eq:betamatch} 
\end{equation} 
Under this matching, $M_{\rm eff}^2 = \frac{1}{8\pi G_N}$ so $M_{\rm eff}$ coincides with the physical reduced Planck mass.

\subsection{Ricci-flat protection through quadratic curvature order}
\label{sec:ricciFlatProtection}

The reduction in Eq.~\eqref{eq:GBreduce} has a background-level consequence independent of the Minkowski expansion. We consider a four-dimensional metric satisfying
\begin{equation}
R_{\mu\nu}=0.
\label{eq:genericRicciFlat}
\end{equation}
The $n=1$ logarithmic contribution is proportional to the Einstein tensor and vanishes on Eq.~\eqref{eq:genericRicciFlat}. The $n=2$ contribution is the Euler density plus $5R_{\mu\nu}R^{\mu\nu}$. The Euler density has no local bulk Euler--Lagrange tensor in four dimensions, while
\begin{align}
H^{(R_{\rho\sigma}^{2})}_{\mu\nu}={}&2R_{\mu\rho\nu\sigma}R^{\rho\sigma}-\frac12g_{\mu\nu}R_{\rho\sigma}R^{\rho\sigma}+\Box R_{\mu\nu}
\notag\\
&+\frac12g_{\mu\nu}\Box R-\nabla_\mu\nabla_\nu R
\label{eq:RicciSquaredExactVariationRF}
\end{align}
vanishes identically on every Ricci-flat metric. This is a standard structural property of four-dimensional quadratic gravity \cite{Stelle:1977ry,Bueno:2016ypa}. In the $R^2$--$C^2$ basis the same conclusion follows because four-dimensional Einstein metrics have vanishing Bach tensor. The GfE-specific content is not the protection itself, but the fact that the logarithm fixes the complete quadratic combination with no independent coupling at this order and then fixes $\Tr_{\Lambda^2}(\mathcal C^3)$ as the next invariant in its particular curvature series.

On a Ricci-flat geometry the scalar and one-form curvature blocks vanish and the two-form block is the Weyl-curvature endomorphism $\mathcal C$. The first logarithmic term that can provide a generic correction to the background is therefore
\begin{equation}
S_{\rm RF}^{(3)}
=\frac{\beta^3}{3\mpl^4}
\int\dd^4x\sqrt{-g}\,
\Tr_{\Lambda^2}(\mathcal C^3).
\label{eq:RicciFlatCubicWeyl}
\end{equation}
This yields the GfE hierarchy
\begin{equation}
\begin{aligned}
\text{generic metric:}&\quad \mathbb R^2\text{ corrections},\\
R_{\mu\nu}=0:&\quad \mathcal C^3\text{ first possible}.
\end{aligned}
\label{eq:RicciFlatHierarchy}
\end{equation}
while the square-zero Ricci-flat subclass discussed in Sec.~\ref{sec:ppwave} is exact at the level of the full background equations. Thus every Ricci-flat Einstein solution is protected through the complete curvature-quadratic order, but a generic member of this class need not solve the full logarithmic theory. The covariant reduction has now fixed both the weak-field Hessian and the first curvature invariant that can provide a generic departure from general relativity on a Ricci-flat background. We next vary the reduced action about Minkowski spacetime and derive the linear field equations in the conventions established above.

For a diagonalizable Ricci-flat background, Eqs.~\eqref{eq:diagonalRicciFlatLog} and \eqref{eq:diagonalRicciFlatExpansion} provide the corresponding eigenvalue representation. The nonzero quadratic density does not alter the conclusion above, because background evaluation and covariant variation are different operations.

A recent symmetry-reduced treatment of spherically symmetric GfE vacua reports corrections to the Schwarzschild metric at the first relative order in the curvature coupling~\cite{Thattarampilly:2026wsw}. That analysis employs a different normalization of the induced curvature term and restricts the flattened curvature operator to a diagonal form before deriving the reduced equations. It is therefore not manifestly equivalent to the covariant metric variation considered here. For the covariant four-dimensional metric-only action, Eq.~\eqref{eq:GBreduce} implies that the complete curvature-quadratic Euler--Lagrange tensor vanishes on every Ricci-flat metric. Consequently, a generic correction to Schwarzschild cannot arise from the curvature-quadratic term and first becomes possible through the cubic Weyl invariant in Eq.~\eqref{eq:RicciFlatCubicWeyl}. Reconciling the two variational procedures is an important question for future work.

\section{Linearized field equations}
\label{sec:lineareq}

The covariant reduction in Sec.~\ref{sec:quadratic} supplies the complete local bulk Minkowski Hessian. We now vary the reduced action explicitly, both to fix signs and normalizations and to prepare the source and polarization calculations.

\subsection{Linearized curvatures}

We remind the reader that indices in this section are moved with $\eta_{\mu\nu}$. At first order, we have
\begin{align}
R^{(1)}_{\mu\nu\rho\sigma}
&=\frac12\left(
\partial_{\rho}\partial_{\nu}h_{\mu\sigma}
+\partial_{\sigma}\partial_{\mu}h_{\nu\rho}
-\partial_{\sigma}\partial_{\nu}h_{\mu\rho}
-\partial_{\rho}\partial_{\mu}h_{\nu\sigma}
\right),
\label{eq:linriemann}\\
R^{(1)}_{\mu\nu}
&=\frac12\left(
\partial_{\rho}\partial_{\mu}h^{\rho}{}_{\nu}
+\partial_{\rho}\partial_{\nu}h^{\rho}{}_{\mu}
-\Box h_{\mu\nu}
-\partial_{\mu}\partial_{\nu}h
\right),
\label{eq:linricci}\\
R^{(1)}
&=\partial_{\mu}\partial_{\nu}h^{\mu\nu}-\Box h,
\label{eq:linscalar}\\
G^{(1)}_{\mu\nu}
&=R^{(1)}_{\mu\nu}-\frac12\eta_{\mu\nu}R^{(1)}.
\label{eq:linEinstein}
\end{align}
They satisfy
\begin{equation}
\partial^{\mu}G^{(1)}_{\mu\nu}=0,
\qquad
\partial^{\mu}R^{(1)}_{\mu\nu}=\frac12\partial_{\nu}R^{(1)}.
\label{eq:linBianchi}
\end{equation}

\subsection{Variation of the curvature-squared action}

For a general four-dimensional bulk action
\begin{equation}
S=\int \dd^{4}x\sqrt{-g}
\left(AR+\alpha R^{2}+B R_{\mu\nu}R^{\mu\nu}\right),
\end{equation}
the linearized vacuum equation is the standard one for local quadratic gravity \cite{Stelle:1977ry,Hindawi:1995an,Salvio:2018crh}:
\begin{equation}
(A+B\Box)G^{(1)}_{\mu\nu}
+(2\alpha+B)
(\eta_{\mu\nu}\Box-\partial_{\mu}\partial_{\nu})R^{(1)}=0.
\label{eq:generalQGlin}
\end{equation}
A derivation is given in Appendix~\ref{app:variation}. For GfE, $\alpha=0$ and the common factor $\beta/\mpl^{4}$ can be divided out. One obtains then
\begin{equation}
3G^{(1)}_{\mu\nu}
+\frac{5\beta}{2}
\left[
\Box G^{(1)}_{\mu\nu}
+(\eta_{\mu\nu}\Box-\partial_{\mu}\partial_{\nu})R^{(1)}
\right]=0.
\label{eq:GfElinear}
\end{equation}
The divergence of Eq.~\eqref{eq:GfElinear} vanishes identically by Eq.~\eqref{eq:linBianchi}, as required by linearized diffeomorphism invariance.

For the quadratic action $A R+B R_{\mu\nu}R^{\mu\nu}$, the established flat-space quadratic-gravity spectrum gives
\begin{equation}
m_0^2=\frac{A}{2B},
\qquad
m_2^2=-\frac{A}{B}.
\label{eq:standardMassFormulas}
\end{equation}
The trace of Eq.~\eqref{eq:GfElinear} provides a direct convention check:
\begin{equation}
-3R^{(1)}+5\beta\Box R^{(1)}=0,
\end{equation}
or
\begin{equation}
(\Box-m_{0}^{2})R^{(1)}=0,
\qquad m_{0}^{2}=\frac{3}{5\beta}.
\label{eq:scalartrace}
\end{equation}
This is precisely the scalar formula in Eq.~\eqref{eq:standardMassFormulas} evaluated at the GfE coefficients. For $\beta>0$, the trace mode is non-tachyonic.

\section{Spin decomposition and pole structure}
\label{sec:spectrum}

The sector decomposition reviewed in this section is the standard flat-space content of local quadratic gravity. We include it to establish the signs and normalizations induced by the GfE coefficient choice and to identify the resulting instability. The linear equation is transverse by construction, so its propagating content is most transparent in the conserved spin-2 and scalar sectors.

\subsection{Massless and additional spin-2 sectors}

Projecting onto the four-dimensionally transverse, traceless spin-2 part $h^{(2)}_{\mu\nu}$, we have
\begin{equation}
\partial^{\mu}h^{(2)}_{\mu\nu}=0,
\qquad
\eta^{\mu\nu}h^{(2)}_{\mu\nu}=0.
\end{equation}
Then
\begin{equation}
R^{(1)}=0,
\qquad
G^{(1)}_{\mu\nu}=-\frac12\Box h^{(2)}_{\mu\nu},
\end{equation}
and Eq.~\eqref{eq:GfElinear} becomes
\begin{equation}
\Box\left(3+\frac{5\beta}{2}\Box\right)h^{(2)}_{\mu\nu}=0.
\end{equation}
Equivalently,
\begin{equation}
\Box(\Box+\mu_{2}^{2})h^{(2)}_{\mu\nu}=0,
\qquad
\mu_{2}^{2}=\frac{6}{5\beta}.
\label{eq:spin2factor}
\end{equation}
There are two branches:
\begin{align}
\Box h^{(2)}_{\mu\nu}&=0,
\label{eq:masslessbranch}\\
(\Box+\mu_{2}^{2})h^{(2)}_{\mu\nu}&=0.
\label{eq:extrabranch}
\end{align}
In the Klein--Gordon convention $(\Box-m^{2})\phi=0$, the second branch has
\begin{equation}
m_{2}^{2}=-\mu_{2}^{2}=-\frac{6}{5\beta}.
\label{eq:m2}
\end{equation}
Together with Eq.~\eqref{eq:scalartrace}, this is the standard pair of masses in Eq.~\eqref{eq:standardMassFormulas} after substitution of the GfE coefficients $A$ and $B$. The locked relation between them, rather than the existence of the two sectors, is the model-specific result.
Thus the positive $\beta$ assumed by GfE makes the additional spin-2 pole tachyonic. For a Fourier mode $e^{\ii(\bm{k}\cdot\bm{x}-\omega t)}$,
\begin{equation}
\omega^{2}=\bm{k}^{2}-\mu_{2}^{2}.
\label{eq:tachdisp}
\end{equation}
Modes with $|\bm{k}|<\mu_{2}$ grow or decay exponentially. The maximum linear growth rate occurs at $\bm{k}=0$ and is
\begin{equation}
\Gamma_{\rm max}=\mu_{2}=\sqrt{\frac{6}{5\beta}}.
\label{eq:growth}
\end{equation}
If Eq.~\eqref{eq:betamatch} is imposed, this scale is Planckian; that may place the instability beyond the range of a classical continuum description, but it does not remove it from the formal linear spectrum of the metric-only logarithmic action.

\subsection{Scalar representative}

A convenient representative of the scalar sector is a conformal perturbation
\begin{equation}
h^{(0)}_{\mu\nu}=\eta_{\mu\nu}\psi.
\label{eq:conformalh}
\end{equation}
The standard Barnes--Rivers spin-zero representative is written with the transverse scalar projector and therefore contains an explicitly momentum-dependent longitudinal term. For the nonzero scalar pole, $p^{2}\neq0$, the difference between that representative and Eq.~\eqref{eq:conformalh} is a linearized gauge transformation of the form $\partial_{(\mu}\xi_{\nu)}$, together with a harmless normalization of $\psi$. The linearized Riemann tensor on a flat background is invariant under such transformations. Consequently, the detector response computed below is independent of this convenient gauge choice.
For this perturbation,
\begin{align}
R^{(1)}_{\mu\nu}
&=-\partial_{\mu}\partial_{\nu}\psi
-\frac12\eta_{\mu\nu}\Box\psi,
\label{eq:scalarRicci}\\
R^{(1)}&=-3\Box\psi,
\label{eq:scalarR}\\
G^{(1)}_{\mu\nu}
&=(\eta_{\mu\nu}\Box-\partial_{\mu}\partial_{\nu})\psi.
\label{eq:scalarG}
\end{align}
Substitution into Eq.~\eqref{eq:GfElinear} gives
\begin{equation}
(3-5\beta\Box)
(\eta_{\mu\nu}\Box-\partial_{\mu}\partial_{\nu})\psi=0.
\end{equation}
After discarding pure-gauge or zero-curvature pieces, the scalar mode obeys
\begin{equation}
(\Box-m_{0}^{2})\psi=0,
\qquad m_{0}^{2}=\frac{3}{5\beta}.
\label{eq:psieq}
\end{equation}
The trace curvature and scalar amplitude are related by
\begin{equation}
R^{(1)}=-3m_{0}^{2}\psi,
\qquad
\psi=-\frac{R^{(1)}}{3m_{0}^{2}}
=-\frac{5\beta}{9}R^{(1)}.
\label{eq:Rpsi}
\end{equation}

\subsection{Mass relation, Einstein normalization, and sign obstruction}

The two nonzero pole masses satisfy
\begin{equation}
m_{2}^{2}=-2m_{0}^{2},
\label{eq:massrelation}
\end{equation}
and at the level of mass squared alone
\begin{equation}
\beta>0 : \quad m_{0}^{2}>0,\qquad m_{2}^{2}<0,
\end{equation}
\begin{equation}
\beta<0\ \text{(formal continuation)} : \quad m_{0}^{2}<0,\qquad m_{2}^{2}>0.
\end{equation}

The second line lies outside the parameter domain of the foundational GfE proposal, which assumes $\beta>0$ \cite{Bianconi:2024aju}; it is included only to diagnose the analytic metric action under a formal sign continuation. Equation~\eqref{eq:GfECouplingRatio} also shows that this continuation reverses the Einstein--Hilbert kinetic term and the sign of $G_{\rm eff}$. Relative to a conventionally normalized positive-energy matter sector, the massless graviton would then have the wrong absolute sign. It is therefore not a viable alternative in which the tachyon simply migrates from spin two to spin zero.

The relevance of the GfE coefficient locking is now immediate. Positive Einstein normalization requires $\beta>0$. For this sign, the standard quadratic-gravity scalar pole is non-tachyonic, whereas the additional spin-2 pole has negative mass squared and residue opposite to the massless graviton. Changing the sign of $\beta$ makes the spin-2 mass squared positive but simultaneously makes the scalar tachyonic and reverses the Einstein term. Thus no point on the GfE coupling ray has conventional Einstein normalization together with non-tachyonic nonzero scalar and spin-2 sectors. This statement concerns the unrestricted analytic metric formulation; an order-reduced EFT or a distinct constrained phase-space prescription is an additional theory choice.

The implication is stronger than the statement that an additional pole is present. In the unrestricted analytic metric formulation, Minkowski spacetime is an exact solution but not a stable perturbative vacuum when the spin-2 branch in Eq.~\eqref{eq:extrabranch} is admitted in the initial data. This affects its interpretation as a ground state, as the asymptotic geometry of isolated systems, and as the background used to define weak-field particle states. The obstruction is not removed by infinitesimal matter fluctuations about $\bar\Phi=0$, because Eq.~\eqref{eq:fullMatterOperator} is quadratic in the topological matter field. It may nevertheless be reorganized on a genuinely matter-supported background, for which the coupled Hessian must be recomputed rather than inferred from the vacuum equation.

\subsection{Diagonal-curvature subcase of the Minkowski spectrum}
\label{sec:diagonalSpectrum}

Minkowski spacetime has $\bar{\mathbb R}=0$, so the background curvature operator is trivially diagonal in every basis. If diagonality is imposed only on the background, nothing changes in the perturbative calculation: the first-order curvature $\delta\mathbb R=\mathbb R^{(1)}[h]$ remains unrestricted, and the complete Hessian and pole structure derived above are recovered.

The situation changes if the curvature perturbation itself is required to be diagonalizable or diagonal in a fixed bivector basis. The scalar representative in Eq.~\eqref{eq:conformalh} provides an example that is compatible with diagonalization. For a plane wave propagating in the $z$-direction and satisfying $\omega^2-k^2=m_0^2$, factor out the scalar wave amplitude from the linearized two-form curvature operator. The remaining $6\times6$ matrix has spectrum
\begin{equation}
\sigma(\widehat{\mathcal R}^{(1)}_0) = \left\{ 0,0,0,-m_0^2,-m_0^2,-m_0^2 \right\},
\label{eq:scalarRiemannEigenvalues}
\end{equation}
and is diagonalizable for $m_0^2\neq0$. Thus the scalar pole survives in the diagonalizable subcase with exactly the same mass $m_0^2=3/(5\beta)$.

By contrast, a nonzero massless transverse-traceless plane wave has type-N linearized curvature. Its two-form curvature endomorphism satisfies
\begin{equation}
\left(\mathcal R_{\rm TT}^{(1)}\right)^2=0,
\qquad
\mathcal R_{\rm TT}^{(1)}\neq0,
\qquad
\sigma(\mathcal R_{\rm TT}^{(1)})=\{0\}.
\label{eq:TTNilpotentCurvature}
\end{equation}

It is therefore nilpotent and non-diagonalizable. A strict diagonalizable-curvature restriction would force $\mathcal R_{\rm TT}^{(1)}=0$ and would exclude, rather than modify, the nontrivial massless GR wave branch. The explicit $6\times6$ scalar and TT representatives are given in Appendix~\ref{app:diagonalRepresentatives}.

The additional spin--2 pole is a property of the full covariant Minkowski Hessian. Whenever a particular perturbation admits a diagonalizable curvature representative, evaluating the trace-log in an eigenbasis leaves the invariant Hessian coefficients unchanged. Its dispersion relation therefore remains $s=-\mu_2^2$. The diagonalizable eigenvalue representation neither generates an additional pole nor removes the tachyonic one. The situation is different when diagonality is imposed as a restriction on the allowed perturbations. A fixed diagonal ansatz may project out some of the five directions contained in the massive spin--2 projector, but this does not modify the pole structure of the full theory. It only reduces the class of perturbations on which that structure is tested. Such a restricted calculation therefore cannot replace the covariant residue analysis or the corresponding degree-of-freedom count.

Three logically distinct operations should therefore be distinguished:
\begin{enumerate}
\item
For diagonalizable curvature blocks, choosing an eigenbasis is merely an equivalent representation of the covariant trace-log. The invariant quadratic coefficients, pole masses, and residues remain unchanged.

\item
Imposing a diagonal background after varying the full covariant action is a legitimate specialization of the resulting field equations. Unless an additional condition is introduced, the perturbations about that background remain unrestricted.

\item
Imposing a fixed diagonal-curvature ansatz directly in the action before variation defines a reduced variational problem. It discards off-diagonal curvature variations and probes only a restricted subsector of the full spectrum. In particular, such a truncation excludes nonzero type-N massless waves and square-zero pp-wave curvature operators, which are intrinsically non-diagonalizable.
\end{enumerate}

\subsection{Standard quadratic-gravity inverse and GfE convention checks}
\label{sec:propagator}
We use the standard Barnes--Rivers decomposition to invert the vacuum quadratic operator independently in its conserved spin--2 and scalar sectors \cite{Rivers:1964nfl,Barnes:1965ylk,VanNieuwenhuizen:1973fi, Stelle:1976gc}. The projector definitions, Hessian eigenvalues, and partial-fraction decomposition are collected in Appendix~\ref{app:projectorDetails}. With the conventions
\begin{equation}
S_{\rm GfE}^{(2)}
=
\frac12\int hKh,
\qquad
\Pi_{\rm alg}
=
\mathrm{i}K^{-1},
\qquad
g_{\mu\nu}
=
\eta_{\mu\nu}+h_{\mu\nu},
\label{eq:hessianconvention}
\end{equation}
the gauge-independent algebraic inverse on the conserved-source subspace is
\begin{align}
\Pi_{\rm alg}(s)
=
\frac{2\mathrm{i}\mpl^{4}}{3\beta}
\bigg[
&\frac{P^{(2)}-\tfrac12P^{(0s)}}{s}
-\frac{P^{(2)}}{s+\mu_{2}^{2}}
\notag\\
&+\frac{\tfrac12P^{(0s)}}{s-m_{0}^{2}}
\bigg].
\label{eq:propagator}
\end{align}
Here $\Pi_{\rm alg}$ is the momentum-space inverse of the vacuum Hessian before a Green-function prescription has been selected. Its poles and algebraic residues are properties of the metric-only quadratic operator and do not require the introduction of dynamical matter.

At momenta small compared with the two nonzero pole scales, Eq.~\eqref{eq:propagator} reduces to
\begin{equation}
\Pi_{\rm alg}^{\rm GR}(s)
=
\frac{4\mathrm{i}}{M_{\rm eff}^{2}}
\frac{P^{(2)}-\tfrac12P^{(0s)}}{s},
\label{eq:GRpropagatorcheck}
\end{equation}
which fixes the normalization of the massless pole for the unrescaled perturbation $h_{\mu\nu}$.

The partial-fraction decomposition also makes the relative residue signs explicit. The scalar pole has the same algebraic residue sign as the massless graviton, whereas the additional spin--2 pole has the opposite sign:
\begin{equation}
\operatorname{sgn}
\left(
\operatorname{Res}_{2,0},
\operatorname{Res}_{0},
\operatorname{Res}_{2,m}
\right)
=
\left(
\operatorname{sgn}\beta,
\operatorname{sgn}\beta,
-\operatorname{sgn}\beta
\right).
\label{eq:absoluteresidues}
\end{equation}
For $\beta>0$, the additional spin--2 pole lies at $s=-\mu_2^2$ and therefore combines an opposite residue with a tachyonic dispersion relation. The conventional expression ``spin--2 ghost'' refers to the relative residue sign; the branch is not a stable massive particle.

Equation~\eqref{eq:propagator} is an algebraic inverse rather than a complete Green function. For
\begin{equation}
|\mathbf k|<\mu_2,
\end{equation}
the additional spin--2 branch has the frequency roots
\begin{equation}
\omega_\pm=\pm\mathrm{i}\sqrt{\mu_2^2-\mathbf k^2}.
\label{eq:tachyonicfrequencypoles}
\end{equation}
A retarded, advanced, Feynman, or initial-value prescription is additional physical input, but it does not change the pole locations or their algebraic residue signs.

For the sole purpose of saturating the inverse and checking its response normalization, we introduce an auxiliary external conserved source,
\begin{equation}
S_{\rm src}=\frac12\int\mathrm d^4x\,h_{\mu\nu}T^{\mu\nu},
\qquad
\partial_\mu T^{\mu\nu}=0.
\label{eq:sourcecoupling}
\end{equation}
The tensor $T_{\mu\nu}$ is not a dynamical GfE matter field and is not identified with the topological matter operator $\widetilde M$. It is a nondynamical bookkeeping source used to probe the already-derived vacuum Hessian. Consequently, Eq.~\eqref{eq:sourcecoupling} does not modify the background condition $|\bar\Phi\rangle=0$, the vacuum Hessian, or its pole structure.

With this source, the linear response equation is
\begin{widetext}
\begin{equation}
3G_{\mu\nu}^{(1)}
+\frac{5\beta}{2}
\left[
\Box G_{\mu\nu}^{(1)}
+
\left(
\eta_{\mu\nu}\Box
-\partial_\mu\partial_\nu
\right)R^{(1)}
\right]
=
\frac{\mpl^4}{2\beta}T_{\mu\nu}.
\label{eq:sourcedLinearEquation}
\end{equation}
\end{widetext}
Its trace gives
\begin{equation}
\left(
\Box-m_0^2
\right)R^{(1)}
=
\frac{\mpl^4}{10\beta^2}T.
\label{eq:sourcedScalarEquation}
\end{equation}
Thus a traceful external source probes the scalar pole, whereas a
traceless source does not couple directly to it at linear order.

\begin{table*}[t]
\centering
\caption{Linearized vacuum spectrum of the metric-only GfE action around Minkowski spacetime. The Barnes--Rivers rank is the algebraic rank for $p^2\neq0$ and is analytically continued when discussing the massless residue; the last column gives the on-shell propagating interpretation after gauge equivalence and constraints. Residues are quoted relative to a positive-energy probe after imposing $\beta>0$.}
\label{tab:spectrum}
\begin{ruledtabular}
\small
\begin{tabular}{lcccc}
Sector & Pole & BR rank & Residue & On-shell content\\
\hline
Massless spin 2 & $\Box h^{(2)}_{\mu\nu}=0$ & $5$ & positive & two helicities: plus/cross\\
Scalar & $(\Box-m_{0}^{2})\psi=0$ & $1$ & positive & one breathing--longitudinal mode\\
Additional spin 2 & $(\Box+\mu_{2}^{2})h^{(2)}_{\mu\nu}=0$ & $5$ & negative & tachyonic rank-five branch; no stable particle\\
\end{tabular}
\end{ruledtabular}
\begin{flushleft}
Here $m_{0}^{2}=3/(5\beta)$ and $\mu_{2}^{2}=6/(5\beta)$, so $m_{2}^{2}=-\mu_{2}^{2}$. For completeness, a formal continuation to $\beta<0$ multiplies these signs by $\operatorname{sgn}\beta$; that continuation lies outside the parameter domain of the foundational proposal.
\end{flushleft}
\end{table*}

\section{Gravitational-wave solutions and detector polarizations}
\label{sec:polarizations}

Having fixed the poles, we translate each branch into wave solutions and gauge-invariant tidal observables. The purpose is to specialize standard polarization results to the GfE masses and sign conventions.

\subsection{The exact linear GR branch}

Any linearized vacuum solution of Einstein gravity satisfies
\begin{equation}
G^{(1)}_{\mu\nu}=0.
\end{equation}
Its trace implies $R^{(1)}=0$. Therefore every such solution automatically satisfies Eq.~\eqref{eq:GfElinear}. It follows that every linearized Ricci-flat GR perturbation around Minkowski spacetime solves the complete linearized equations of the analytic continuum logarithmic action. In Fourier space this includes each null mode, $p^{2}=0$, within that continuum description. This statement is algebraically stronger than a low-frequency truncation, but it is not a claim of effective-theory validity at arbitrarily large null frequency. A derivative cutoff may restrict which null Fourier solutions are admissible. Nor does the statement imply that a generic finite-amplitude Ricci-flat GR spacetime solves the full nonlinear logarithmic theory.

For a wave propagating in the $+z$ direction, we can impose the usual spatial transverse-traceless gauge,
\begin{equation}
h_{0\mu}^{\rm TT}=0,
\qquad
\partial^{i}h_{ij}^{\rm TT}=0,
\qquad
h^{{\rm TT}\,i}{}_{i}=0.
\end{equation}
Then
\begin{equation}
h_{ij}^{\rm TT}(t,z)
=h_{+}(u)e^{+}_{ij}+h_{\times}(u)e^{\times}_{ij},
\qquad u=t-z,
\label{eq:TTwave}
\end{equation}
with
\begin{equation}
e^{+}_{ij}=
\begin{pmatrix}1&0&0\\0&-1&0\\0&0&0\end{pmatrix},
\qquad
e^{\times}_{ij}=
\begin{pmatrix}0&1&0\\1&0&0\\0&0&0\end{pmatrix}.
\end{equation}
The dispersion relation is
\begin{equation}
\omega^{2}=k^{2},
\end{equation}
so both phase and group velocities are unity.

The measurable tidal tensor is
\begin{equation}
\cE_{ij}\equiv R^{(1)}_{0i0j}
=-\frac12\ddot h^{\rm TT}_{ij}.
\label{eq:GRtidal}
\end{equation}
Geodesic separation vectors obey
\begin{equation}
\ddot\xi^{i}=-\cE^{i}{}_{j}\xi^{j},
\end{equation}
which yields the standard plus and cross distortions.

\subsection{Scalar breathing and longitudinal response}

We take the scalar representative in Eq.~\eqref{eq:conformalh} and a plane wave
\begin{equation}
\psi=\psi_{0}e^{\ii(kz-\omega t)},
\qquad
\omega^{2}=k^{2}+m_{0}^{2}.
\label{eq:scalarplane}
\end{equation}
We use
\begin{equation}
R^{(1)}_{0i0j}=\frac12\partial_{i}\partial_{j}\psi-\frac12\delta_{ij}\partial_{0}^{2}\psi,
\label{eq:scalarR0i0j}
\end{equation}
and obtain
\begin{align}
\cE_{xx}=\cE_{yy}&=\frac12\omega^{2}\psi,
\label{eq:breathing}\\
\cE_{zz}&=\frac12(\omega^{2}-k^{2})\psi=\frac12m_{0}^{2}\psi,
\label{eq:longitudinal}\\
\cE_{xy}=\cE_{xz}=\cE_{yz}&=0.
\end{align}
An explicit component derivation of Eqs.~\eqref{eq:breathing}--\eqref{eq:longitudinal} is provided in Appendix~\ref{app:tidal}.

The previous result is the standard mixed breathing--longitudinal response of a massive scalar metric mode \cite{Eardley:1973br,Eardley:1973zuo,Alves:2023rxs}, with the scalar mass fixed here by the GfE coefficient relation. The ratio of the two tidal components is
\begin{equation}
\frac{\cE_{zz}}{\cE_{xx}}=\frac{m_{0}^{2}}{\omega^{2}}.
\label{eq:polratio}
\end{equation}
At high frequency, $\omega\gg m_{0}$, the longitudinal response is suppressed and the mode approaches a predominantly transverse breathing pattern. Near threshold, the longitudinal contribution is comparable to the transverse contribution. The GfE-specific input in this standard response is $m_0^2=3/(5\beta)$.

\subsection{Additional spin-2 sector}

The pole associated with Eq.~\eqref{eq:extrabranch} lies in the rank-five Barnes--Rivers spin-2 sector familiar from the standard linear analysis of symmetric tensor fields and quadratic gravity \cite{Fierz:1939ix,Rivers:1964nfl,Barnes:1965ylk,VanNieuwenhuizen:1973fi,Hindawi:1995an}. Its tensor, vector-like, and scalar-like decomposition is algebraically analogous to the continuation of a massive spin-2 field. Because $m_{2}^{2}<0$ for $\beta>0$, however, its momentum orbit is tachyonic and it is not a physical five-polarization massive-particle representation. For $k<\mu_{2}$, the amplitudes are exponential rather than oscillatory; for $k>\mu_{2}$, they oscillate according to Eq.~\eqref{eq:tachdisp}. Furthermore, Eq.~\eqref{eq:propagator} gives this pole a negative residue. Its observable tidal response may be decomposed directly into tensor, vector-like, and scalar-like components, but the standard null-wave $E(2)$ classification does not apply: Eq.~\eqref{eq:tachdisp} gives $p^{2}=\mu_{2}^{2}>0$, and the long-wavelength modes are not oscillatory waves at all.

\section{The \texorpdfstring{$G$}{G}-field dressing of gravitational waves}
\label{sec:Gfield}

The metric spectrum follows from the reduced logarithmic action. The auxiliary formulation nevertheless assigns a nontrivial $G$-field response to the same waves, which provides an equivalent bookkeeping description on the algebraic branch used here rather than a new propagating branch by itself.

The auxiliary formulation of Ref.~\cite{Bianconi:2024aju} introduces a direct-sum field $\widetilde\cG$ obeying in vacuum
\begin{equation}
\widetilde\cG^{-1}
=\widetilde I-\beta\mathbb R.
\label{eq:Gconstraint}
\end{equation}
We write
\begin{equation}
\widetilde\cG=\widetilde I+\widetilde q.
\end{equation}
To first order,
\begin{equation}
\widetilde q=\beta\mathbb R^{(1)}.
\label{eq:qR}
\end{equation}

The algebraic equation is naturally an equation for mixed form-space endomorphisms. In the component formulas below we lower the one-form and two-form output indices with the Minkowski background metric; for example, $q_{(1)\mu\nu}\equiv\eta_{\nu\rho}q_{(1)\mu}{}^{\rho}$ and $q_{(2)\mu\nu\rho\sigma}\equiv\eta_{\rho\alpha}\eta_{\sigma\beta}q_{(2)\mu\nu}{}^{\alpha\beta}$. With this convention we have
\begin{align}
q_{(0)}&=\beta R^{(1)},
\label{eq:q0}\\
q_{(1)\mu\nu}&=\beta R^{(1)}_{\mu\nu},
\label{eq:q1}\\
q_{(2)\mu\nu\rho\sigma}&=\beta R^{(1)}_{\mu\nu\rho\sigma}.
\label{eq:q2}
\end{align}
For the ordinary massless GR wave,
\begin{equation}
R^{(1)}=0,
\qquad
R^{(1)}_{\mu\nu}=0,
\qquad
R^{(1)}_{\mu\nu\rho\sigma}\neq0,
\end{equation}
and thus
\begin{equation}
q_{(0)}=0,
\qquad q_{(1)}=0,
\qquad q_{(2)}=\beta R^{(1)}_{\mu\nu\rho\sigma}\neq0.
\label{eq:GRqdress}
\end{equation}
A standard GR gravitational wave is thus accompanied by a nontrivial two-form $G$-field dressing even though its metric propagation remains exactly that of linearized GR.

The emergent cosmological term in the auxiliary formulation is
\begin{equation}
\Lambda_{\cG}=\frac{1}{2\beta}\TrF\left(\widetilde\cG-\widetilde I-\operatorname{Log}\widetilde\cG\right).
\label{eq:Lambdag}
\end{equation}
For $\widetilde\cG=\widetilde I+\widetilde q$ this reduces to
\begin{equation}
\Lambda_{\cG}=\frac{1}{4\beta}\TrF(\widetilde q^{2})+\order(\widetilde q^{3}).
\label{eq:Lambdaquad}
\end{equation}
It has no linear effect on waves about Minkowski. It should not automatically be identified with the Isaacson energy density: for type-N plane waves, scalar curvature contractions such as $R_{\mu\nu\rho\sigma}R^{\mu\nu\rho\sigma}$ can vanish even while the tidal field is nonzero.

\section{Independent check from the auxiliary \texorpdfstring{$G$}{G}-field equations}
\label{sec:auxcheck}

We now verify the coefficient of the four-derivative operator directly in the auxiliary formulation. This is a consistency check of the GfE reduction, not a rederivation of the general auxiliary-field formulation of quadratic gravity. In the enlarged variational formulation the metric and $G$-field are varied independently; only after linearization do we impose the algebraic $G$-field equation and eliminate its perturbation. The three form-space blocks, the two-form trace normalization, and the resulting coefficient are displayed explicitly below and in Appendix~\ref{app:auxDerivative}. Commutator terms vanish at this order because the background direct-sum operator is the identity.

We use the foundational conventions of Ref.~\cite{Bianconi:2024aju} and write the metric field equation in the schematic vacuum form
\begin{equation}
R^{\cG}_{(\mu\nu)}
-\frac12g_{\mu\nu}(R_{\cG}-2\Lambda_{\cG})
+D_{(\mu\nu)}=0.
\label{eq:auxmetric}
\end{equation}
At linear order about
$g_{\mu\nu}=\eta_{\mu\nu}$ and $\widetilde\cG=\widetilde I$, the algebraic curvature dressing in $R^{\cG}_{\mu\nu}$ reduces to
\begin{equation}
R^{\cG(1)}_{\mu\nu}=3R^{(1)}_{\mu\nu},
\qquad
R_{\cG}^{(1)}=3R^{(1)},
\qquad
\Lambda_{\cG}^{(1)}=0.
\label{eq:Rglinear}
\end{equation}
Hence the non-derivative part is $3G^{(1)}_{\mu\nu}$.

The derivative term $D_{\mu\nu}$ is linear in second derivatives of the three form-degree blocks of $\widetilde q$. Appendix~\ref{app:auxDerivative} displays the linearized block operator before the curvature constraint is substituted and derives each contribution. We then use Eqs.~\eqref{eq:q0}--\eqref{eq:q2} to obtain
\begin{align}
D^{[0]}_{\mu\nu}
&=\beta(\eta_{\mu\nu}\Box-\partial_{\mu}\partial_{\nu})R^{(1)},
\label{eq:D0}\\
D^{[1]}_{\mu\nu}
&=\beta\left[\frac12\Box R^{(1)}_{\mu\nu}-\frac12\partial_{\mu}\partial_{\nu}R^{(1)}+\frac14\eta_{\mu\nu}\Box R^{(1)}
\right],
\label{eq:D1}\\
D^{[2]}_{\mu\nu}&=\beta\left[2\Box R^{(1)}_{\mu\nu}-\partial_{\mu}\partial_{\nu}R^{(1)}
\right].
\label{eq:D2}
\end{align}
The two-form result uses
\begin{align}
\partial^{\eta}\partial^{\rho}R^{(1)}_{\mu\rho\nu\eta}
&=\Box R^{(1)}_{\mu\nu}
-\frac12\partial_{\mu}\partial_{\nu}R^{(1)},
\label{eq:bianchiT1}\\
\partial^{\rho}\partial^{\eta}R^{(1)}_{\eta\mu\rho\nu}
&=\Box R^{(1)}_{\mu\nu}
-\frac12\partial_{\mu}\partial_{\nu}R^{(1)}.
\label{eq:bianchiT2}
\end{align}
Adding Eqs.~\eqref{eq:D0}--\eqref{eq:D2} gives the total linear perturbation,
\begin{align}
\delta D_{\mu\nu}
&=\frac{5\beta}{2}\Box R^{(1)}_{\mu\nu}
+\frac{5\beta}{4}\eta_{\mu\nu}\Box R^{(1)}
-\frac{5\beta}{2}\partial_{\mu}\partial_{\nu}R^{(1)}
\nonumber\\
&=\frac{5\beta}{2}
\left[
\Box G^{(1)}_{\mu\nu}
+(\eta_{\mu\nu}\Box-\partial_{\mu}\partial_{\nu})R^{(1)}
\right].
\label{eq:Dtotal}
\end{align}
Substitution into Eq.~\eqref{eq:auxmetric} reproduces Eq.~\eqref{eq:GfElinear} exactly. This is an independent check of the factor $5/2$ and of the relative scalar derivative term.

The use of auxiliary fields to expose the scalar and spin-2 content of fourth-order gravity is standard \cite{Whitt:1984pd,Hindawi:1995an}. The present check is narrower: it verifies the coefficient and sign map of the foundational GfE $G$-field equation.

The check also clarifies the phase-space issue. If the $G$-field equation \eqref{eq:Gconstraint} is enforced and $\widetilde q$ is eliminated, the metric equation is fourth order and has the spectrum in Table~\ref{tab:spectrum}. Declaring $\widetilde\cG$ to be a physically independent field while simultaneously changing the allowed initial data or boundary conditions could define a different canonical theory, as already noted in Ref.~\cite{Bianconi:2024aju}. Such an additional prescription would need to be stated explicitly before claiming that any of the poles derived here are absent.

\section{Square-zero Ricci-flat curvature and exact pp-waves}
\label{sec:ppwave}

The persistence of Ricci-flat pp-waves in broad classes of higher-curvature theories is well known from the universality literature \cite{Pravda:2002us,Hervik:2013cla,Hervik:2017sdr,Bueno:2016ypa}. We now isolate what is more specific to GfE: the mixed curvature endomorphism is square-zero, so the matrix logarithm and resolvent reduce exactly to finite expressions, and the local bulk Hessian has a finite curvature-order cutoff.

\subsection{Ricci-flat Brinkmann waves}

We consider the Brinkmann metric~\cite{Brinkmann:1925fr,Podolsky:2014lpa},
\begin{equation}
\dd s^{2}=-2\,\dd u\,\dd v+\dd x^{2}+\dd y^{2} +H(u,x,y)\,\dd u^{2}.
\label{eq:ppmetric}
\end{equation}
Its only independent nonzero curvature components are
\begin{equation}
R_{uiuj}=-\frac12\partial_i\partial_j H,
\qquad i,j\in\{x,y\},
\label{eq:ppRiemann}
\end{equation}
and
\begin{equation}
R_{uu}=-\frac12(\partial_x^2+\partial_y^2)H.
\end{equation}
The metric is Ricci-flat and hence a vacuum solution of Einstein gravity, if and only if
\begin{equation}
(\partial_x^2+\partial_y^2)H=0.
\label{eq:ppvac}
\end{equation}
This class contains the usual plane-wave subclass
\begin{equation}
H(u,x,y)=A_{+}(u)(x^{2}-y^{2})+2A_{\times}(u)xy,
\label{eq:ppPlusCross}
\end{equation}
with arbitrary smooth plus and cross waveforms $A_{+}$ and $A_{\times}$. More general harmonic functions $H(u,x,y)$ describe the broader Ricci-flat pp-wave family.

A nonzero Riemann component in Eq.~\eqref{eq:ppRiemann} carries a covariant $u$ in each antisymmetric index pair. Raising the second pair converts that $u$ into a $v$, whereas every component with a covariant $v$ in the first pair vanishes. The two-form curvature endomorphism therefore satisfies
\begin{equation}
\mathcal R_{\mu\nu}{}^{\alpha\beta}\mathcal R_{\alpha\beta}{}^{\rho\sigma}=0.
\label{eq:nilpotent}
\end{equation}
For a Ricci-flat pp-wave the scalar and one-form blocks vanish, so the complete direct-sum operator is square-zero,
\begin{equation}
\mathbb R^2=0.
\label{eq:fullnilpotent}
\end{equation}

Whenever the curvature is nonzero, Eq.~\eqref{eq:fullnilpotent} also proves that the operator is not diagonalizable: a diagonal square-zero matrix would vanish identically. Thus the pp-wave sector is absent from a strict diagonal-curvature truncation even though it is fully admitted by the foundational matrix formulation. This is consistent with the vanishing-scalar-invariant and universal-spacetime properties of Ricci-flat type-N waves \cite{Pravda:2002us,Hervik:2013cla,Hervik:2017sdr}. Under analytic matrix functional calculus, Eq.~\eqref{eq:fullnilpotent} gives
\begin{equation}
\operatorname{Log}(\widetilde I-\beta\mathbb R)=-\beta\mathbb R
\label{eq:nilpotentlog}
\end{equation}
exactly. The local field equations nevertheless require the first variation, not only the value of the Lagrangian. The $n=1$ term vanishes by Ricci flatness. The $n=2$ term vanishes by the Ricci-flat protection established in Sec.~\ref{sec:ricciFlatProtection}. For every $n\geq3$,
\begin{equation}
\delta\Tr_F(\mathbb R^n)
=n\Tr_F(\mathbb R^{n-1}\delta\mathbb R)=0,
\label{eq:nilpotentVariation}
\end{equation}
while the measure variation also vanishes because $\Tr_F(\mathbb R^n)=0$. It follows that every smooth Ricci-flat Brinkmann pp-wave solves the local bulk Euler--Lagrange equations of the metric-only logarithmic action, provided the logarithm is taken on the analytic branch connected to $\operatorname{Log}\widetilde I=0$, the Euler density is treated as topological, and boundary contributions are absent or cancelled by an appropriate variational prescription. The statement is specifically that the higher logarithmic terms have zero value and zero \emph{first variation} on the square-zero background. It does not imply that their second or higher variations vanish.

A direct matrix-logarithm calculation makes the finite Hessian structure transparent. We define $\mathbb A=\widetilde I-\beta\mathbb R$. Then
\begin{align}
\delta[-\Tr_F\operatorname{Log}\mathbb A]
&=\beta\Tr_F(\mathbb A^{-1}\delta\mathbb R),
\label{eq:directLogFirstVariation}\\
\delta^2[-\Tr_F\operatorname{Log}\mathbb A]
&=\beta\Tr_F(\mathbb A^{-1}\delta^2\mathbb R)
\notag\\
&\quad+\beta^2\Tr_F(\mathbb A^{-1}\delta\mathbb R\,
\mathbb A^{-1}\delta\mathbb R).
\label{eq:directLogSecondVariation}
\end{align}
On the square-zero background, $\bar{\mathbb R}^{2}=0$ and
\begin{equation}
\bar{\mathbb A}^{-1}=\widetilde I+\beta\bar{\mathbb R}.
\end{equation}
Substitution into Eq.~\eqref{eq:directLogSecondVariation} gives
\begin{align}
\delta^2[-\Tr_F\operatorname{Log}\mathbb A]
={}&\beta\Tr_F(\delta^2\mathbb R)
+\beta^2\Tr_F(\bar{\mathbb R}\delta^2\mathbb R)
\notag\\
&+\beta^2\Tr_F[(\delta\mathbb R)^2]
+2\beta^3\Tr_F[\bar{\mathbb R}(\delta\mathbb R)^2]
\notag\\
&+\beta^4\Tr_F(\bar{\mathbb R}\delta\mathbb R\,
\bar{\mathbb R}\delta\mathbb R).
\label{eq:directLogHessianFinite}
\end{align}
Thus the matrix-function part of the Hessian contains no power above $\beta^4$. The measure variations do not restore higher powers because the background density and its first variation vanish for all logarithmic terms with $n\ge3$. The same cutoff can be checked term by term. For a one-parameter perturbation of a noncommuting curvature endomorphism,
\begin{align}
\delta^2\Tr_F(\mathbb R^n)
={}&n\Tr_F\!\left(\bar{\mathbb R}^{\,n-1}\delta^2\mathbb R\right)
\notag\\
&+n\sum_{k=0}^{n-2}\Tr_F\!\left(
\bar{\mathbb R}^{\,k}\delta\mathbb R\,
\bar{\mathbb R}^{\,n-2-k}\delta\mathbb R
\right).
\label{eq:noncommutingSecondVariation}
\end{align}
Even when $\bar{\mathbb R}^{2}=0$, separated insertions can survive. The cubic term contains
\begin{equation}
\delta^2\Tr_F(\mathbb R^3)
=6\Tr_F\!\left[\bar{\mathbb R}(\delta\mathbb R)^2\right],
\end{equation}
up to the term involving $\bar{\mathbb R}^{2}\delta^2\mathbb R$, which vanishes, while the quartic term can contain
\begin{equation}
4\Tr_F\!\left(\bar{\mathbb R}\delta\mathbb R\bar{\mathbb R}\delta\mathbb R\right).
\end{equation}
Nilpotence nevertheless imposes an exact upper bound. In every term of the cyclic sum in Eq.~\eqref{eq:noncommutingSecondVariation}, survival requires both background powers to be at most one,
\begin{equation}
k\le1,
\qquad n-2-k\le1,
\end{equation}
which implies $n\le4$. The first term in Eq.~\eqref{eq:noncommutingSecondVariation} already vanishes for $n\ge3$. Moreover, for $n\ge3$ the background density and its first variation vanish, so the second variation of the measure and the measure--first-variation cross term cannot restore any $n\ge5$ contribution. Thus
\begin{equation}
\left.\delta^2\!\left[\sqrt{-g}\,\Tr_F(\mathbb R^n)\right]
\right|_{\bar{\mathbb R}^{2}=0}=0,
\qquad n\ge5.
\label{eq:ppHessianTruncation}
\end{equation}
The complete local bulk Hessian of the logarithmic action about a square-zero background therefore truncates exactly after the quartic curvature term. The $n=3$ and $n=4$ terms may modify the perturbation operator, so exactness of the background does not imply a GR-like pp-wave stability spectrum. More generally, the $q$th variation about a square-zero operator can receive curvature-power contributions only for $n\le2q$, because the $q$ varied insertions can separate at most $q$ background factors. The local background result still does not establish finiteness of the total action, a global boundary-value problem, global hyperbolicity, or admissibility in a separately postulated positive entropy-operator domain.

\subsection{Arbitrary profile amplitude}

The exactness of this sector does not rely on a weak-curvature or small-amplitude expansion. We take $H$ to be any smooth harmonic profile satisfying Eq.~\eqref{eq:ppvac} and rescale it by an arbitrary finite real parameter,
\begin{equation}
H(u,x,y)\longrightarrow \lambda H(u,x,y).
\end{equation}

For a Ricci-flat pp-wave, the curvature endomorphism is square-zero,
\begin{equation}
\mathbb R^2=0.
\label{eq:ppSquareZeroReminder}
\end{equation}

Consequently, an arbitrary rescaling of the wave amplitude preserves
nilpotency,
\begin{equation}
(\lambda\mathbb R)^2=0,
\end{equation}
and the matrix-logarithm series therefore terminates after its linear
term:
\begin{align}
\operatorname{Log}
(\widetilde I-\beta\lambda\mathbb R)
&=
-\sum_{n=1}^{\infty}
\frac{(\beta\lambda\mathbb R)^n}{n}
\notag\\
&=
-\beta\lambda\mathbb R.
\label{eq:ppArbitraryAmplitudeLog}
\end{align}

This identity is exact and does not rely on $\lVert\beta\lambda\mathbb R\rVert\ll1$. Indeed, because $\lambda\mathbb R$ is nilpotent, all eigenvalues of $\widetilde I-\beta\lambda\mathbb R$ are equal to unity, so the principal matrix logarithm remains well defined for every finite $\lambda$. The result therefore remains valid even when a chosen matrix norm satisfies
\begin{equation}
\lVert\beta\lambda\mathbb R\rVert\gg1.
\end{equation}

Thus the analytic matrix logarithm imposes no local small-amplitude or small-curvature restriction on this square-zero family. Global regularity, boundary conditions, and any independent ultraviolet regime of validity of the continuum theory remain separate questions.

\subsection{Nilpotent curvature and spectral degeneracy of the local density}

A nontrivial pp-wave has $\mathbb R\neq0$, but nilpotency implies
\begin{equation}
\Tr_F(\mathbb R^n)=0
\qquad (n\geq1),
\label{eq:nilpotentTraceInvariants}
\end{equation}
 together with
\begin{equation}
\det(\widetilde I-\beta\mathbb R)=1,
\qquad
\Tr_F\operatorname{Log}(\widetilde I-\beta\mathbb R)=0.
\label{eq:ppSpectralDegeneracy}
\end{equation}
More generally, if $f$ is analytic near the origin, square-zero nilpotence gives
\begin{equation}
f(\mathbb R)=f(0)\widetilde I+f'(0)\mathbb R,
\qquad
\Tr_F f(\mathbb R)=11f(0).
\label{eq:ppAnalyticTraceDegeneracy}
\end{equation}
The right-hand side is exactly the Minkowski value $\Tr_F[f(0)\widetilde I]$. Thus every positive-power curvature trace vanishes, every normalized analytic trace functional with $f(0)=0$ vanishes, and every general analytic trace functional takes the same value as in Minkowski spacetime. The distinction resides in the nilpotent Jordan part of the curvature operator, which is not encoded by its eigenvalues or by analytic traces.

The same degeneracy is visible in the auxiliary description. Equation~\eqref{eq:Gconstraint} gives
\begin{equation}
\widetilde\cG^{-1}=\widetilde I-\beta\mathbb R,
\qquad
\widetilde\cG=\widetilde I+\beta\mathbb R,
\qquad
\operatorname{Log}\widetilde\cG=\beta\mathbb R.
\label{eq:ppAuxiliaryExact}
\end{equation}
Consequently,
\begin{equation}
\widetilde\cG-\widetilde I=\beta\mathbb R\neq0,
\qquad
\widetilde\cG-\widetilde I-\operatorname{Log}\widetilde\cG=0,
\qquad
\Lambda_{\cG}=0.
\label{eq:ppLambdaZero}
\end{equation}
The metric universality of Ricci-flat type-N pp-waves in broad higher-curvature theories is known \cite{Hervik:2013cla,Hervik:2017sdr,Bueno:2016ypa}. The GfE-specific statement is the exact auxiliary realization in Eqs.~\eqref{eq:ppAuxiliaryExact}--\eqref{eq:ppLambdaZero}: the two-form $G$-field dressing and the Riemann tensor are nontrivial even though $\Lambda_{\cG}=0$, all positive-power curvature traces vanish, and every analytic trace functional takes its Minkowski value according to Eq.~\eqref{eq:ppAnalyticTraceDegeneracy}. This should not be described without qualification as zero physical entropy: the positive-operator interpretation motivating GfE and the Lorentzian analytic continuation need not have the same admissible domain. It does show that eigenvalue data and analytic trace invariants alone do not distinguish Minkowski curvature from a nilpotent radiative curvature operator.

\section{Averaged flux of the pure massless TT branch}
\label{sec:energy}

The pole analysis identifies a healthy massless eigenspace for $\beta>0$. We close the wave discussion with a restricted statement: on the isolated massless transverse-traceless eigenspace, the averaged translation current derived from the quadratic action has the standard GR normalization.

The energy statement needed here is narrower than a full nonlinear Isaacson calculation. In spatial transverse-traceless gauge, $R^{(1)}=0$ and $R_{ij}^{(1)}=-\tfrac12\Box h_{ij}^{\rm TT}$. The quadratic TT action obtained from Eq.~\eqref{eq:reducedaction} is
\begin{align}
S_{\rm TT}^{(2)}=\int\dd^4x\bigg[&-\frac{M_{\rm eff}^2}{8}
\partial_\lambda h_{ij}^{\rm TT}\partial^\lambda h_{\rm TT}^{ij}\notag\\
&+\frac{B}{4}(\Box h_{ij}^{\rm TT})(\Box h_{\rm TT}^{ij})\bigg].
\label{eq:TTquadraticaction}
\end{align}
On the isolated massless eigenspace,
\begin{equation}
\Box h_{ij}^{\rm TT}=0,
\label{eq:masslessBranchCondition}
\end{equation}
all $B$-dependent terms vanish from the reduced symplectic current and from the translation current. A convenient symmetric conserved representative is therefore
\begin{equation}
\tau_{\mu\nu}^{(h)}
=\frac{M_{\rm eff}^{2}}{4}\left[
\partial_\mu h_{ij}^{\rm TT}\partial_\nu h_{\rm TT}^{ij}
-\frac12\eta_{\mu\nu}\partial_\lambda h_{ij}^{\rm TT}\partial^\lambda h_{\rm TT}^{ij}\right].
\label{eq:reducedEnergyTensor}
\end{equation}
For a short-wavelength average or a monochromatic null wave, integration by parts and Eq.~\eqref{eq:masslessBranchCondition} one can remove the trace term, giving
\begin{align}
\left\langle\tau_{\mu\nu}^{(h)}\right\rangle
&=\frac{M_{\rm eff}^{2}}{4}
\left\langle\partial_\mu h_{ij}^{\rm TT}\partial_\nu h_{\rm TT}^{ij}\right\rangle\\
&=\frac{1}{32\pi G_{\rm eff}}
\left\langle\partial_\mu h_{ij}^{\rm TT}\partial_\nu h_{\rm TT}^{ij}\right\rangle.
\label{eq:reducedFlux}
\end{align}
This agrees with the standard high-frequency GR normalization of Isaacson for the averaged flux of the \emph{pure massless TT branch} \cite{Isaacson:1968hbi,Isaacson:1968zza}. It neither establishes positivity of the Hamiltonian of the complete fourth-order theory nor constitutes a complete second-order Isaacson analysis of the full logarithmic field equations, to which curvature-cubic and higher terms may contribute. Such an analysis requires a second-order background-field expansion of the complete logarithmic equations before branch restriction. Appendix~\ref{app:energyDetails} records the reduced-current check and the assumptions for general TT packets.

\section{Scope beyond the vacuum Minkowski branch}
\label{sec:matterScope}

The conserved source in Eq.~\eqref{eq:sourcecoupling} is a conventional probe and does not reproduce the complete dynamical matter sector. In the foundational GfE theory, matter and curvature enter the same logarithm through Eqs.~\eqref{eq:fullMatterOperator} and \eqref{eq:fullInducedMetric}. Moreover, $\widetilde M$ is itself metric-dependent: the Hodge--Dirac operator, the form-space contractions, and the nonminimal factor $\xi R$ all vary with the metric. A nonzero background $\bar\Phi$ can therefore modify both the background equations and the coupled metric--matter--$G$ Hessian. It may change effective kinetic coefficients, generate metric--matter mixing, and shift local characteristic scales and pole locations. This possibility is physically relevant to the vacuum obstruction found above, but it is not by itself a stabilization mechanism. In particular, the presence of the term $\xi R|\Phi\rangle\langle\Phi|$ does not establish that the opposite-residue spin-2 branch is removed. That conclusion would require the complete constrained Hessian about a self-consistent background with $\bar\Phi\neq0$, including all metric, matter, and $G$-field perturbations. Stabilization could arise only if the resulting kinetic operator shifts the branch to a stable region, renders it nondynamical through an additional constraint, or excludes it from the physical phase space.

Moreover, curved vacuum backgrounds also differ qualitatively from Minkowski. For a generic $\bar{\mathbb R}\neq0$, Eq.~\eqref{eq:noncommutingSecondVariation} contains all cyclic orderings of background and perturbed curvature operators; cyclicity does not collapse them to one ordering unless $\bar{\mathbb R}$ commutes with $\delta\mathbb R$. A generic curved-background local bulk Hessian can therefore receive contributions from every logarithmic power, and the fixed Minkowski map to $A R+B R_{\mu\nu}^2$ cannot be transplanted unchanged. The square-zero pp-wave is an exceptional finite case rather than an example of the generic statement: Eq.~\eqref{eq:ppHessianTruncation} removes every $n\ge5$ contribution, leaving only the curvature orders $n=1,2,3,4$ in its exact local bulk Hessian.

Cosmological gravitational waves are consequently a separate calculation rather than a direct extension of the Minkowski pole analysis. On an evolving FLRW background there is no global pole mass in the flat-space sense. One must first select a background solution of the same covariant formulation, derive the gauge- and constraint-reduced tensor system, and study its kinetic and gradient matrices, local WKB characteristics, and primordial spectrum. Matter-supported cosmology may change the coefficients and mix the metric with topological matter and auxiliary variables, but a partial two-derivative parametrization would not establish the physical spectrum. The present stability conclusion is therefore restricted to the metric-only analytic vacuum branch.

\section{Physical interpretation and observational implications}
\label{sec:discussion}

\subsection{What is unchanged from GR}
As is customary when analyzing modified theories of gravity, an important question is how the theory reduces to GR and which results remain unchanged. The massless GfE branch is exactly the linearized GR branch of the continuum equations. It has
\begin{equation}
\omega^{2}=k^{2},
\end{equation}
two tensor polarizations, the standard weak-wave tidal response, and the standard positive flat-space quadratic energy normalization when $\beta>0$. There is no frequency-dependent correction to its Minkowski dispersion relation. A calculation restricted consistently to this eigenspace reproduces ordinary vacuum propagation on flat spacetime.

The $G$-field nevertheless records the wave through its two-form block, Eq.~\eqref{eq:GRqdress}. Thus ``the metric wave behaves as in GR'' does not mean that every field introduced by GfE is trivial.

\subsection{What the diagonal restriction does and does not test}
\label{sec:diagonalInterpretation}

The diagonalizable eigenvalue representation is a computational specialization of the covariant action, not an alternative theory. Equations~\eqref{eq:diagonalLogExpansion}--\eqref{eq:diagonalSecondSpectralSums} show that it reproduces the same quadratic  invariants and therefore the same coefficients and pole locations. It is especially useful for homogeneous, isotropic, or static backgrounds whose mixed curvature operators are diagonal in an adapted basis.

A fixed diagonal-curvature ansatz imposed before variation has a narrower meaning. It tests only perturbations tangent to that reduced configuration space. It cannot establish the absence of a mode whose curvature lies outside the ansatz. The massless TT branch illustrates this point sharply: its nonzero type-N curvature is nilpotent and non-diagonalizable according to Eq.~\eqref{eq:TTNilpotentCurvature}. A diagonal truncation would therefore remove the ordinary GR wave sector itself. The same is true of the exact pp-waves in Sec.~\ref{sec:ppwave}. Consequently, the diagonal sector can be used to evaluate diagonalizable backgrounds and selected perturbations, but not to replace the full covariant stability analysis.

\subsection{Order-reduced vacuum dynamics through quadratic curvature order}
\label{sec:vacuumOrderReduction}

The literal fourth-order continuum equation and its perturbatively order-reduced EFT counterpart have different solution spaces. At the first higher-curvature order, the vacuum metric equation is Eq.~\eqref{eq:GfElinear}. About Minkowski spacetime the leading on-shell equation is
\begin{equation}
G_{\mu\nu}^{(1)}=0,
\qquad R^{(1)}=0.
\label{eq:leadingVacuumEFT}
\end{equation}
We substitute Eq.~\eqref{eq:leadingVacuumEFT} into the term proportional to $\beta$. The complete correction then vanishes, so the order-reduced linear vacuum equation remains
\begin{equation}
G_{\mu\nu}^{(1)}=0
\label{eq:orderReducedVacuumGR}
\end{equation}
through quadratic curvature order. More generally, about a Ricci-flat GR background $\bar g_{\mu\nu}$, the same order-reduced conclusion holds for a leading perturbation satisfying the on-shell conditions $\delta R_{\mu\nu}[\bar g;h]=0$ and $\delta R[\bar g;h]=0$. It does not follow that the unreduced curved-background Hessian is the Einstein Hessian: although Eq.~\eqref{eq:RicciSquaredExactVariationRF} vanishes on the background, its linearization need not vanish for an arbitrary off-shell perturbation. Vacuum deviations in an on-shell order-reduced expansion can first arise from the cubic Weyl sector in Eq.~\eqref{eq:RicciFlatCubicWeyl}.

This result shows the difference between two interpretations. In the literal analytic continuum theory, the exact local bulk Minkowski Hessian contains the scalar and tachyonic spin-2 poles. After perturbative order reduction, leading on-shell vacuum perturbations about Minkowski---or more generally about a Ricci-flat GR background---obey the GR propagation equations through curvature-quadratic order. Order reduction is an additional prescription for the admissible solution space; it does not erase the poles from the unreduced Hessian.

\subsection{Locked pole scales at fixed microscopic length}
\label{sec:noDecoupling}

The parameter $\beta$ controls both the normalization of the Einstein term and the masses of the additional poles. Indeed, using Eqs.~\eqref{eq:GfECouplingRatio}, \eqref{eq:scalartrace}, and
\eqref{eq:spin2factor}, we find
\begin{equation}
m_0^2 M_{\rm eff}^2\mpl^4=\frac{18}{5},
\qquad
\mu_2^2 M_{\rm eff}^2\mpl^4=\frac{36}{5}.
\label{eq:lockedPoleProducts}
\end{equation}
These combinations are independent of $\beta$. Consequently, at fixed $\mpl$, the pole scales cannot be sent to infinity by varying $\beta$ while keeping the Einstein normalization fixed. In particular, Eq.~\eqref{eq:Geff} shows that fixing $G_{\rm eff}$ and $\mpl$ already fixes $\beta$, whereas the formal limit $\beta\to0$ gives
\begin{equation}
M_{\rm eff}^2\to0,
\qquad
G_{\rm eff}\to\infty.
\end{equation}
Thus $\beta\to0$ is not a GR decoupling limit of the theory at fixed
$\mpl$.

If $\mpl$ were instead treated as an independent microscopic
length, rather than as the physical Planck length, we can formally
consider the simultaneous scaling
\begin{equation}
\mpl\to0,
\qquad
\beta=\frac{M_{\rm eff}^2\mpl^4}{6}\to0,
\label{eq:formalTwoParameterDecoupling}
\end{equation}
with $M_{\rm eff}$ held fixed. In that scaling,
\begin{equation}
m_0^2
=
\frac{18}{5M_{\rm eff}^2\mpl^4},
\qquad
\mu_2^2
=
\frac{36}{5M_{\rm eff}^2\mpl^4},
\end{equation}
and both nonzero pole scales diverge. This is only a formal
two-parameter limit and is unavailable when $\mpl$ is identified
with the ordinary Planck length.

With the standard definition
\begin{equation}
\mpl^2=G_N
\end{equation}
in units $\hbar=c=1$, matching the low-energy coupling to the
measured Newton constant,
\begin{equation}
G_{\rm eff}=G_N,
\end{equation}
fixes
\begin{equation}
\beta=\frac{\mpl^2}{48\pi}.
\end{equation}
The nonzero pole scales are then locked to the Planck scale:
\begin{equation}
m_0^2G_N=\frac{144\pi}{5},
\qquad
\mu_2^2G_N=\frac{288\pi}{5}.
\label{eq:PlanckLockedMasses}
\end{equation}
They are therefore not independently tunable parameters once the
Einstein normalization and the standard meaning of $\mpl$ have
been fixed.

\subsection{Fundamental-continuum and EFT interpretations of the nonzero poles}

The nonzero scales are
\begin{equation}
m_{0}^{2}=\frac{3}{5\beta},
\qquad
\mu_{2}^{2}=\frac{6}{5\beta},
\qquad m_{2}^{2}=-\mu_{2}^{2}.
\end{equation}
For the required sign $\beta>0$, the scalar obeys
\begin{equation}
\omega^{2}=k^{2}+m_{0}^{2}.
\end{equation}
If $m_{0}$ is Planckian, a low-frequency astrophysical source cannot emit this mode as an on-shell propagating wave. Its observational suppression is a genuine heavy-particle threshold.

The additional rank-five spin-2 projector branch is qualitatively different:
\begin{equation}
\omega^{2}=k^{2}-\mu_{2}^{2}.
\end{equation}
For $k<\mu_{2}$,
\begin{equation}
h_{\mu\nu}\propto
\exp\!\left[\pm\sqrt{\mu_{2}^{2}-k^{2}}\,t\right].
\label{eq:tachgrowth}
\end{equation}

Long-wavelength perturbations therefore grow rather than facing a production threshold. For astrophysical $k\ll\mu_{2}$, the growth rate approaches $\mu_{2}$; it is not suppressed by the low source frequency. If the logarithmic continuum action is fundamental, Minkowski spacetime is unstable on the timescale $\mu_{2}^{-1}$.

If the theory is instead an effective field theory with a cutoff $\Lambda\ll\mu_{2}$, the runaway solutions may be excluded because their characteristic time derivatives lie outside the admissible derivative regime. The explicit order-reduced result in Sec.~\ref{sec:vacuumOrderReduction} shows that leading on-shell vacuum perturbations then remain governed by the Einstein equation through quadratic curvature order \cite{Simon:1990ic,Parker:1993dk}. This is not equivalent to saying that the tachyon is merely too heavy to produce: it is a prescription selecting the low-energy solution branch.

With the conventional matching in Eq.~\eqref{eq:betamatch},
\begin{equation}
m_{0}=\sqrt{\frac{144\pi}{5}}\,\mpl^{-1},
\qquad
\mu_{2}=\sqrt{\frac{288\pi}{5}}\,\mpl^{-1}.
\end{equation}
The interpretation is therefore bifurcated. If the analytic metric action is fundamental, these are genuine poles of the exact local bulk Minkowski Hessian and the tachyon implies a Planck-scale instability timescale. If the logarithmic expansion is used only as a low-energy EFT with cutoff well below $\beta^{-1/2}$, neither nonzero pole should be treated as an asymptotic state; the unstable branch is excluded only after an explicit order-reduction or branch-selection prescription. The coefficient and mass-relation calculation is valid in either reading, but the physical state space is not the same.

\subsection{Metric-only versus independent-\texorpdfstring{$G$}{G} interpretations}

The original logarithmic action is higher derivative when written solely in terms of $g_{\mu\nu}$. The auxiliary formulation rewrites the equations as second order in $g_{\mu\nu}$ and $\widetilde\cG$, but its algebraic field equation ties $\widetilde\cG$ to curvature. Enforcing that relation and eliminating $\widetilde\cG$ reproduces the fourth-order equation and the spectrum above.

A different spectrum is possible only if the independent-$G$ formulation is supplemented by a genuinely different phase-space prescription, constraints, contour choice, or boundary condition. Such a prescription may be physically motivated, but it is extra information; auxiliary rewriting alone does not prove ghost freedom. Recent work has developed a thermodynamic Hamiltonian treatment of homogeneous and isotropic GfE cosmologies \cite{Bianconi:2025awa}, but we are not aware of a complete nonlinear ADM constraint analysis of the unrestricted theory that determines its physical propagating degrees of freedom. Accordingly, no claim about the independent-$G$ spectrum is made beyond the explicit elimination check performed here.

\section{Conclusions}
\label{sec:conclusions}

The first input of this work is specific to Gravity from Entropy. We used the foundational form-space normalization to evaluate the direct-sum traces explicitly and obtained Eqs.~\eqref{eq:trace1} and \eqref{eq:trace2}. Together with the four-dimensional Euler identity \eqref{eq:GBreduce}, these results fix the curvature-quadratic coefficients to
\begin{equation}
A=\frac{3\beta}{\mpl^4},
\qquad
B=\frac{5\beta^2}{2\mpl^4}.
\end{equation}
Because $\mathbb R=\order(h)$ about Minkowski spacetime, Eq.~\eqref{eq:MinkowskiHessianIdentity} shows that these coefficients determine the complete local bulk Minkowski Hessian. This direct-sum calculation, including the two-form normalization, is the essential model-specific step.

The diagonalizable subcase provides an independent representation of the same result. We write the Ricci and Riemann blocks in eigenbases, which converts the trace-log into Eq.~\eqref{eq:diagonalEigenvalueLog}; its expansion then reproduces the invariant traces and coefficients without modification. This equivalence must be distinguished from imposing a fixed diagonal-curvature ansatz before variation. The latter gives only a reduced Hessian and can exclude physical sectors: the massive scalar curvature is diagonalizable, whereas a nonzero massless TT wave has square-zero, non-diagonalizable curvature as shown in Eq.~\eqref{eq:TTNilpotentCurvature}. Thus diagonality does not cure or alter the covariant pole structure; it either provides an equivalent eigenvalue representation or restricts the perturbations being tested.

The subsequent flat-space mode decomposition is the established spectrum of four-dimensional quadratic gravity. We apply the standard formulas \eqref{eq:standardMassFormulas} to the GfE coefficients and obtain
\begin{equation}
m_0^2=\frac{3}{5\beta},
\qquad
m_2^2=-\frac{6}{5\beta}=-2m_0^2.
\end{equation}
The standard Barnes--Rivers inversion likewise gives the propagator \eqref{eq:propagator}: the scalar residue has the graviton sign, whereas the additional four-derivative spin-2 pole has the opposite residue. The new implication is produced by the GfE coefficient locking. Positive Einstein normalization requires $\beta>0$, which makes the scalar non-tachyonic but places the opposite-residue spin-2 projector branch on the tachyonic side. Reversing $\beta$ makes the scalar tachyonic and also reverses the Einstein term. No point on the GfE coupling ray therefore combines conventional Einstein normalization with non-tachyonic nonzero scalar and spin-2 sectors. This is the Minkowski-stability obstruction for the unrestricted analytic metric formulation.

The auxiliary-$G$ calculation supplies an independent GfE coefficient check. Varying the metric and $G$-field independently, linearizing the three form-degree blocks, and then imposing the algebraic relation \eqref{eq:vacuumGfieldEquation} gives the contributions \eqref{eq:D0}--\eqref{eq:D2}. Their sum is Eq.~\eqref{eq:Dtotal}, reproducing the four-derivative operator in Eq.~\eqref{eq:GfElinear}. The general auxiliary-field interpretation of quadratic gravity is standard; the role of this calculation is to verify the GfE normalization and the factor $5/2$. A different physical mode count would require a genuinely different phase-space, constraint, contour, or boundary prescription.

At the nonlinear background level, the vanishing of curvature-quadratic field equations on four-dimensional Ricci-flat metrics is also a standard property of quadratic gravity. In GfE, the nontrivial result is that the direct-sum logarithm selects precisely this protected quadratic combination and then fixes Eq.~\eqref{eq:RicciFlatCubicWeyl} as the first possible generic correction on a Ricci-flat background. The resulting logarithmic hierarchy is summarized in Eq.~\eqref{eq:RicciFlatHierarchy}. This background protection does not imply that the unreduced Hessian about an arbitrary curved Ricci-flat metric is the Einstein Hessian.

Ricci-flat pp-waves are known to persist in broad classes of higher-curvature theories. Their nonzero square-zero curvature also demonstrates why the covariant operator formulation is strictly more general than a diagonalizable-curvature restriction. The additional GfE result follows from the square-zero identity \eqref{eq:fullnilpotent}. It makes the logarithm and resolvent finite at arbitrary profile amplitude, yielding exact local bulk background solutions on the analytic branch. More specifically, the direct Hessian calculation \eqref{eq:directLogHessianFinite} contains powers only through $\beta^4$, while the equivalent curvature-power statement \eqref{eq:ppHessianTruncation} removes every logarithmic term with $n\ge5$. Thus exact background solvability does not imply GR perturbations, but it converts the formally infinite local bulk Hessian into a finite calculation involving curvature orders $n=1,2,3,4$.

The standard scalar-polarization analysis, evaluated at the GfE mass, gives the mixed breathing--longitudinal response \eqref{eq:breathing}--\eqref{eq:polratio}. On the isolated massless transverse-traceless eigenspace, the quadratic translation current has the standard GR flux normalization in Eq.~\eqref{eq:reducedFlux}. This restricted statement is not a complete second-order Isaacson analysis of the analytic logarithmic equations and is not a positivity result for the full higher-derivative theory.

The matter sector provides a possible route by which the vacuum spectrum may be reorganized. This is not simply because matter sources curvature, but because $\widetilde M$ in Eq.~\eqref{eq:fullMatterOperator} depends explicitly on the metric, including through the nonminimal coupling $\xi R|\Phi\rangle\langle\Phi|$. On a background with $\bar\Phi\neq0$, this term can alter the pure metric Hessian and produce metric--matter mixing. The present vacuum result therefore does not exclude stabilization on a matter-supported background. At the same time, the curvature dependence of $\widetilde M$ does not affect the pure metric Hessian about $(\bar g_{\mu\nu},\bar\Phi)=(\eta_{\mu\nu},0)$, because $\widetilde M$ is bilinear in $\Phi$. Nor does the presence of the nonminimal coupling alone demonstrate that the spin-2 instability is removed. Establishing such a mechanism requires a self-consistent matter-supported background and an explicit diagonalization of the complete constrained metric--matter--$G$ perturbation operator.

Finally, the interpretation depends on whether the analytic metric action is used as an unrestricted continuum theory or as an effective theory with perturbative order reduction. In the former, the tachyonic spin-2 pole belongs to the exact Minkowski Hessian, so Minkowski spacetime is a solution but not a stable perturbative vacuum when this branch is included in the admissible initial data. In the latter, leading on-shell vacuum perturbations obey the GR propagation equation after the branch selection summarized by Eq.~\eqref{eq:orderReducedVacuumGR}. Order reduction therefore does not change the Hessian; it supplies an additional prescription for selecting its perturbative solutions. The locked products in Eq.~\eqref{eq:lockedPoleProducts} further show that, at fixed $\ell_{\rm P}$, the additional scales cannot be independently decoupled by varying $\beta$ alone. 

The restriction to vacuum does not make this conclusion secondary. Matter is structurally central to GfE because Eqs.~\eqref{eq:fullMatterOperator} and \eqref{eq:fullInducedMetric} place matter and curvature inside the same matrix logarithm. Nevertheless, $\widetilde M=\mathcal O(\Phi^2)$, so infinitesimal matter fluctuations about the state $(\bar g_{\mu\nu},\bar\Phi)=(\eta_{\mu\nu},0)$ do not modify the vacuum gravitational Hessian. The instability found here is therefore a property of the zero-matter Minkowski branch of the full construction, rather than an artifact of omitting a linear matter source. 

This result is not a no-go theorem for every matter-supported realization of GfE. A background with $\bar\Phi\neq0$ can modify the induced operator, the logarithmic domain, the kinetic coefficients, and the mixing among metric, matter, and $G$-field perturbations. Such a resolution must, however, be demonstrated by computing the complete constrained Hessian about a self-consistent matter-supported solution. In particular, it must establish whether the additional spin-2 branch is shifted, removed by a constraint, or excluded from the physical state space, and whether any stabilization persists in regions where the matter background becomes dilute. 

The vacuum analysis therefore supplies a necessary benchmark for the theory. Any viable completion must explain either why the zero-matter Minkowski state is not physically admissible, how a nonzero matter background stabilizes its coupled perturbations, why the auxiliary constraint structure removes the additional branch, or why order reduction is part of the definition of the physical theory. At the same time, the Ricci-flat hierarchy and exact pp-wave sector show that GfE retains a distinguished GR-like class of vacuum geometries. Their coexistence with a non-GR Minkowski spectrum demonstrates that exact background solvability alone is not sufficient to establish perturbative viability. All conclusions remain restricted to the analytic matrix-logarithm domain in Eq.~\eqref{eq:principal-domain}. Boundary and global questions, the complete matter-supported spectrum, generic cosmological backgrounds, and perturbations of the exact pp-wave sector remain open.

\begin{acknowledgments}
 We thank Jess Rutschi for helpful comments. This work was supported by Fundação para a Ciência e a Tecnologia (FCT) through national funds under the research grant UID/04434/2025 (DOI 10.54499/UID/04434/2025).
\end{acknowledgments}

\appendix

\section{Projector and source-normalization details}
\label{app:projectorDetails}

For momentum $p_\mu$, we define
\begin{equation}
\theta_{\mu\nu}=\eta_{\mu\nu}-\frac{p_\mu p_\nu}{p^2},
\qquad
P^{(0s)}_{\mu\nu,\rho\sigma}=\frac13\theta_{\mu\nu}\theta_{\rho\sigma},
\end{equation}
and
\begin{equation}
P^{(2)}_{\mu\nu,\rho\sigma}
=\frac12(\theta_{\mu\rho}\theta_{\nu\sigma}+\theta_{\mu\sigma}\theta_{\nu\rho})
-\frac13\theta_{\mu\nu}\theta_{\rho\sigma}.
\end{equation}
On the conserved-source subspace, the Einstein low-energy Hessian for the unrescaled field is
\begin{equation}
K_{\rm EH}=\frac{M_{\rm eff}^2}{4}sP^{(2)}-\frac{M_{\rm eff}^2}{2}sP^{(0s)}.
\end{equation}
The GfE pole polynomials then give
\begin{align}
K_{\rm cons}={}&\frac{M_{\rm eff}^2}{4}s\left(1+\frac{s}{\mu_2^2}\right)P^{(2)}
-\frac{M_{\rm eff}^2}{2}s\left(1-\frac{s}{m_0^2}\right)P^{(0s)}.
\end{align}
Orthogonality of the projectors immediately yields Eq.~\eqref{eq:propagator}. 

\section{Linearized form-degree derivation of the auxiliary derivative tensor}
\label{app:auxDerivative}

For the flat-background check in Sec.~\ref{sec:auxcheck}, we denote the scalar, one-form, and two-form blocks of $\widetilde q$ by $q_{(0)}$, $q_{(1)\mu\nu}$, and $q_{(2)\mu\nu\rho\sigma}$, with the index-lowering convention stated below Eq.~\eqref{eq:qR}. Before imposing the curvature constraint, the linearized derivative tensor decomposes as
\begin{equation}
\delta D_{\mu\nu}
=\mathscr D^{[0]}_{\mu\nu}
+\mathscr D^{[1]}_{\mu\nu}
+\mathscr D^{[2]}_{\mu\nu},
\label{eq:unreducedDdecomposition}
\end{equation}
where
\begin{align}
\mathscr D^{[0]}_{\mu\nu}
={}&(\eta_{\mu\nu}\Box-\partial_\mu\partial_\nu)q_{(0)},
\label{eq:unreducedD0}\\
\mathscr D^{[1]}_{\mu\nu}
={}&\frac12\Box q_{(1)\mu\nu}
-\partial_\rho\partial_{(\mu}q_{(1)\nu)}{}^\rho
+\frac12\eta_{\mu\nu}
\partial_\rho\partial_\sigma q_{(1)}^{\rho\sigma},
\label{eq:unreducedD1}\\
\mathscr D^{[2]}_{\mu\nu}
={}&\partial^\eta\partial^\rho
q_{(2)\mu\rho\nu\eta}
+\partial^\rho\partial^\eta
q_{(2)\eta\mu\rho\nu}.
\label{eq:unreducedD2}
\end{align}
These formulas are the flat linearization of the form-degree derivative terms in the foundational $G$-field equation \cite{Bianconi:2024aju}; no curvature relation has yet been used.

On the algebraic vacuum branch, Eqs.~\eqref{eq:q0}--\eqref{eq:q2} give $q_{(0)}=\beta R^{(1)}$, $q_{(1)\mu\nu}=\beta R^{(1)}_{\mu\nu}$, and $q_{(2)\mu\nu\rho\sigma}=\beta R^{(1)}_{\mu\nu\rho\sigma}$. The scalar block immediately yields Eq.~\eqref{eq:D0}. For the one-form block, the linearized contracted Bianchi identity implies
\begin{equation}
\partial_\rho q_{(1)\nu}{}^\rho
=\frac{\beta}{2}\partial_\nu R^{(1)},
\qquad
\partial_\rho\partial_\sigma q_{(1)}^{\rho\sigma}
=\frac{\beta}{2}\Box R^{(1)},
\end{equation}
which reduces Eq.~\eqref{eq:unreducedD1} to Eq.~\eqref{eq:D1}. For the two-form block, the linearized differential and contracted Bianchi identities give Eqs.~\eqref{eq:bianchiT1} and \eqref{eq:bianchiT2}; substituting them into Eq.~\eqref{eq:unreducedD2} gives Eq.~\eqref{eq:D2}. Adding the three blocks then produces Eq.~\eqref{eq:Dtotal}. This makes the auxiliary check reproducible without reconstructing the form-degree contractions from the foundational paper.

\section{Reduced TT current and averaging details}
\label{app:energyDetails}

For one TT component, the reduced Lagrangian is
\begin{equation}
\mathcal L=-\frac{M_{\rm eff}^2}{8}(\partial h)^2+\frac{B}{4}(\Box h)^2.
\end{equation}
Its symplectic potential may be chosen as
\begin{equation}
\Theta^\mu=-\frac{M_{\rm eff}^2}{4}(\partial^\mu h)\delta h
+\frac{B}{2}\left[(\Box h)\partial^\mu\delta h-(\partial^\mu\Box h)\delta h\right].
\end{equation}
Every $B$-dependent term in the associated symplectic current vanishes when both perturbations lie on the massless eigenspace $\Box\delta h=0$. The generalized translation current gives Eq.~\eqref{eq:reducedEnergyTensor} after the same restriction. For a general multidirectional packet this tensor is constructed directly from $h_{ij}^{\rm TT}(x)$; a pair of local scalar polarization amplitudes is valid pointwise only for fixed propagation direction. Improvement terms change the local representative by a divergence but not the integrated four-momentum under standard falloff. Finally,
\begin{align}
\left\langle(\partial h_{ij}^{\rm TT})^2\right\rangle
={}&-\left\langle h_{ij}^{\rm TT}\Box h_{\rm TT}^{ij}\right\rangle\notag\\
&+\text{averaged boundary term}=0.
\end{align}
which gives Eq.~\eqref{eq:reducedFlux}.

\section{Variation formulas for quadratic curvature gravity}
\label{app:variation}

We consider
\begin{equation}
S=\int\dd^{4}x\sqrt{-g}
\left(AR+\alpha R^{2}+BR_{\rho\sigma}R^{\rho\sigma}\right).
\end{equation}
The exact metric variations, up to an overall convention for defining the Euler--Lagrange tensor, include
\begin{align}
H^{(R^{2})}_{\mu\nu}
={}&2RR_{\mu\nu}-\frac12g_{\mu\nu}R^{2}
-2\nabla_{\mu}\nabla_{\nu}R
+2g_{\mu\nu}\Box R,
\label{eq:HR2}\\
H^{(R_{\rho\sigma}^{2})}_{\mu\nu}
={}&2R_{\mu\rho\nu\sigma}R^{\rho\sigma}
-\frac12g_{\mu\nu}R_{\rho\sigma}R^{\rho\sigma}
\notag\\
&+\Box R_{\mu\nu}
+\frac12g_{\mu\nu}\Box R
-\nabla_{\mu}\nabla_{\nu}R.
\label{eq:HRic2}
\end{align}
On a flat background, all products of background curvature vanish, leaving
\begin{align}
H^{(R^{2})(1)}_{\mu\nu}
&=2(\eta_{\mu\nu}\Box-\partial_{\mu}\partial_{\nu})R^{(1)},
\label{eq:linHR2}\\
H^{(R_{\rho\sigma}^{2})(1)}_{\mu\nu}
&=\Box R^{(1)}_{\mu\nu}
+\frac12\eta_{\mu\nu}\Box R^{(1)}
-\partial_{\mu}\partial_{\nu}R^{(1)}
\nonumber\\
&=\Box G^{(1)}_{\mu\nu}
+(\eta_{\mu\nu}\Box-\partial_{\mu}\partial_{\nu})R^{(1)}.
\label{eq:linHRic2}
\end{align}
Adding the Einstein term yields Eq.~\eqref{eq:generalQGlin}.

The trace is
\begin{align}
0
&=-AR^{(1)}+(6\alpha+2B)\Box R^{(1)},
\end{align}
so
\begin{equation}
(\Box-m_{0}^{2})R^{(1)}=0,
\qquad
m_{0}^{2}=\frac{A}{2(3\alpha+B)}.
\end{equation}
The transverse spin-2 equation is
\begin{equation}
\Box(A+B\Box)h^{(2)}_{\mu\nu}=0,
\end{equation}
so the nonzero spin-2 mass squared is
\begin{equation}
m_{2}^{2}=-\frac{A}{B}.
\end{equation}
We use Eq.~\eqref{eq:AB} to recover Eqs.~\eqref{eq:scalartrace} and \eqref{eq:m2}.

\section{General scalar- and one-form curvature weights}
\label{app:generalCurvatureWeights}

In this appendix we retain the independent weights $a_0$ and
$a_1$ introduced in Eq.~\eqref{eq:generalTopCurvature}. The
corresponding mixed curvature endomorphism is
\begin{equation}
\mathbb R_{(a_0,a_1)}
=
a_0 R
\oplus
a_1 R_\mu{}^\nu
\oplus
\mathcal R_{\mu\nu}{}^{\rho\sigma},
\label{eq:generalWeightedCurvatureOperator}
\end{equation}
where the normalization of the two-form endomorphism is the one fixed
in Eqs.~\eqref{eq:bivectorop} and
\eqref{eq:twoFormTraceConvention}.

The first direct-sum trace is
\begin{align}
\Tr_F\mathbb R_{(a_0,a_1)}
&=
a_0R
+
a_1\Tr_1(R_\mu{}^\nu)
+
\Tr_2\mathcal R
\nonumber\\
&=
(a_0+a_1+1)R.
\label{eq:generalWeightedFirstTrace}
\end{align}
The second trace is
\begin{align}
\Tr_F\mathbb R_{(a_0,a_1)}^2
={}&
a_0^2R^2
+
a_1^2R_{\mu\nu}R^{\mu\nu}
\nonumber\\
&+
R_{\mu\nu\rho\sigma}R^{\mu\nu\rho\sigma}.
\label{eq:generalWeightedSecondTrace}
\end{align}

The logarithmic action through quadratic curvature order is therefore
\begin{align}
S_{\leq2,\mathrm{curv}}^{(a_0,a_1)}
={}&
\frac{1}{\mpl^4}
\int\dd^4x\sqrt{-g}
\bigg\{
\beta(a_0+a_1+1)R
\nonumber\\
&+
\frac{\beta^2}{2}
\bigg[
a_0^2R^2
+
a_1^2R_{\mu\nu}R^{\mu\nu}
\nonumber\\
&\hspace{2.5cm}
+
R_{\mu\nu\rho\sigma}R^{\mu\nu\rho\sigma}
\bigg]
\bigg\}.
\label{eq:generalWeightedQuadraticAction}
\end{align}
We use the four-dimensional Euler-density identity
\eqref{eq:GB} to rewrite the quadratic combination as
\begin{align}
a_0^2R^2+a_1^2R_{\mu\nu}R^{\mu\nu}+R_{\mu\nu\rho\sigma}R^{\mu\nu\rho\sigma}={}&E_4+(a_0^2-1)R^2
\nonumber\\
&+(a_1^2+4)R_{\mu\nu}R^{\mu\nu}.
\label{eq:generalWeightedGBReduction}
\end{align}
After discarding the Euler density, which does not contribute to the local four-dimensional bulk equations, the local bulk action becomes
\begin{align}
S_{\leq2,\mathrm{curv}}^{(a_0,a_1)}
\doteq
\int\dd^4x\sqrt{-g}
&\left[A_{a_0a_1}R+\alpha_{a_0a_1}R^2 \right.\nonumber\\
&\left.+B_{a_0a_1}R_{\mu\nu}R^{\mu\nu}
\right],
\label{eq:generalWeightedReducedAction}
\end{align}
with
\begin{align}
A_{a_0a_1}
&=
\frac{\beta}{\mpl^4}(a_0+a_1+1),
\label{eq:generalWeightedA}\\
\alpha_{a_0a_1}
&=
\frac{\beta^2}{2\mpl^4}(a_0^2-1),
\label{eq:generalWeightedAlpha}\\
B_{a_0a_1}
&=
\frac{\beta^2}{2\mpl^4}(a_1^2+4).
\label{eq:generalWeightedB}
\end{align}

Substitution into the general linearized equation
\eqref{eq:generalQGlin} gives
\begin{align}
0={}&
\left[
A_{a_0a_1}
+
B_{a_0a_1}\Box
\right]
G_{\mu\nu}^{(1)}
\nonumber\\
&+
\left[
2\alpha_{a_0a_1}
+
B_{a_0a_1}
\right]
(\eta_{\mu\nu}\Box-\partial_\mu\partial_\nu)
R^{(1)}.
\label{eq:generalWeightedLinearEquation}
\end{align}
Its trace is
\begin{equation}
\left[-A_{a_0a_1}+2\left(3\alpha_{a_0a_1}+B_{a_0a_1}\right)\Box\right]R^{(1)}=0.
\label{eq:generalWeightedTraceEquation}
\end{equation}
The scalar pole therefore has mass
\begin{align}
m_0^2(a_0,a_1)&=\frac{A_{a_0a_1}}{2\left(3\alpha_{a_0a_1}+B_{a_0a_1}\right)}
\nonumber\\
&=\frac{a_0+a_1+1}{\beta\left(3a_0^2+a_1^2+1\right)}.
\label{eq:generalWeightedScalarMass}
\end{align}

In the transverse-traceless spin-2 sector,$R^{(1)}=0$ and$G_{\mu\nu}^{(1)}=-\tfrac12\Box h_{\mu\nu}^{(2)}$, so Eq.~\eqref{eq:generalWeightedLinearEquation} reduces to
\begin{equation}
\Box\left(A_{a_0a_1}+B_{a_0a_1}\Box\right)h_{\mu\nu}^{(2)}=0.
\label{eq:generalWeightedSpinTwoEquation}
\end{equation}
The nonzero spin-2 pole is therefore
\begin{align}
m_2^2(a_0,a_1)&=-\frac{A_{a_0a_1}}{B_{a_0a_1}}
\nonumber\\
&=-\frac{2(a_0+a_1+1)}{\beta(a_1^2+4)}.
\label{eq:generalWeightedSpinTwoMass}
\end{align}
The ratio of the two nonzero masses is
\begin{equation}
\frac{m_2^2}{m_0^2}=-2\,\frac{3a_0^2+a_1^2+1}{a_1^2+4}.
\label{eq:generalWeightedMassRatio}
\end{equation}
For the equal-weight choice $a_0=a_1=1$, Eqs.~\eqref{eq:generalWeightedScalarMass} and \eqref{eq:generalWeightedSpinTwoMass} reduce to
\begin{equation}
m_0^2=\frac{3}{5\beta},
\qquad
m_2^2=-\frac{6}{5\beta}=-2m_0^2,
\end{equation}
as obtained in the main text.

For $\beta>0$ and nonnegative weights $a_0,a_1$, we have
\begin{equation}
A_{a_0a_1}>0,
\qquad
m_0^2(a_0,a_1)>0,
\qquad
m_2^2(a_0,a_1)<0.
\label{eq:generalWeightedSignObstruction}
\end{equation}
Thus the tachyonic sign of the additional spin-2 branch is not specific to the equal-weight specialization. The equal-weight choice fixes the particular numerical relation $m_2^2=-2m_0^2$, whereas the incompatibility between conventional Einstein normalization and a non-tachyonic additional spin-2 sector persists throughout the nonnegative $(a_0,a_1)$ parameter domain.

For completeness, the equivalent $R^2$--$C^2$ couplings are
\begin{align}
a_{a_0a_1}&=\alpha_{a_0a_1}+\frac13B_{a_0a_1}
\nonumber\\
&=\frac{\beta^2}{6\mpl^4}\left(3a_0^2+a_1^2+1\right),
\label{eq:generalWeightedRScalarCoupling}\\
b_{a_0a_1}&=\frac12B_{a_0a_1}=\frac{\beta^2}{4\mpl^4}\left(a_1^2+4\right).
\label{eq:generalWeightedWeylCoupling}
\end{align}

\section{Diagonalizable scalar and nilpotent TT curvature representatives}
\label{app:diagonalRepresentatives}

This appendix gives explicit two-form curvature matrices supporting
the statements in Sec.~\ref{sec:diagonalSpectrum}. We use the
independent-pair basis
\begin{equation}
A=(01,02,03,12,13,23).
\label{eq:bivectorBasisDiagonalAppendix}
\end{equation}

For the scalar perturbation
$h_{\mu\nu}=\eta_{\mu\nu}\psi$, we take
$\psi=\psi_0e^{\ii(kz-\omega t)}$ and factor out the common
amplitude $\psi$. The resulting two-form curvature endomorphism is
\begin{equation}
\widehat{\mathcal R}^{(1)}_0
=
\begin{pmatrix}
-\omega^2 & 0 & 0 & 0 & k\omega & 0\\
0 & -\omega^2 & 0 & 0 & 0 & k\omega\\
0 & 0 & k^2-\omega^2 & 0 & 0 & 0\\
0 & 0 & 0 & 0 & 0 & 0\\
-k\omega & 0 & 0 & 0 & k^2 & 0\\
0 & -k\omega & 0 & 0 & 0 & k^2
\end{pmatrix}.
\label{eq:scalarRiemannMatrix}
\end{equation}
Its characteristic polynomial is
\begin{equation}
\chi_0(z)
=
z^3\left[z+\left(\omega^2-k^2\right)\right]^3.
\label{eq:scalarRiemannCharacteristic}
\end{equation}
On the scalar mass shell
$\omega^2-k^2=m_0^2$, this gives
Eq.~\eqref{eq:scalarRiemannEigenvalues}. The geometric and algebraic
multiplicities agree for $m_0^2\neq0$, so the scalar curvature
endomorphism is diagonalizable.

For a massless transverse-traceless wave propagating in the
$z$-direction, we write
\begin{align}
h_{11}^{\rm TT}&=h_+,
&
h_{22}^{\rm TT}&=-h_+,
\notag\\
h_{12}^{\rm TT}&=h_{21}^{\rm TT}=h_\times,
&
\omega&=k.
\end{align}
After factoring out the common plane-wave phase, the corresponding
two-form curvature operator is
\begin{equation}
\mathcal R_{\rm TT}^{(1)}
=
k^2
\begin{pmatrix}
-h_+ & -h_\times & 0 & 0 & h_+ & h_\times\\
-h_\times & h_+ & 0 & 0 & h_\times & -h_+\\
0 & 0 & 0 & 0 & 0 & 0\\
0 & 0 & 0 & 0 & 0 & 0\\
-h_+ & -h_\times & 0 & 0 & h_+ & h_\times\\
-h_\times & h_+ & 0 & 0 & h_\times & -h_+
\end{pmatrix}.
\label{eq:TTRiemannMatrix}
\end{equation}
Direct multiplication gives
\begin{equation}
\left(\mathcal R_{\rm TT}^{(1)}\right)^2=0.
\label{eq:TTRiemannMatrixSquare}
\end{equation}
For a nonzero wave,
$\mathcal R_{\rm TT}^{(1)}\neq0$, while its characteristic
polynomial is $z^6$. The operator is therefore nilpotent of index
two and cannot be diagonalized. A strict diagonalizable-curvature
condition would retain only the trivial zero-curvature member of this
TT family.

\section{Detailed two-form trace normalization}
\label{app:trace}

The bivector metric and its inverse are
\begin{align}
[g_{(2)}]_{\mu\nu\rho\sigma}&=\frac12(g_{\mu\rho}g_{\nu\sigma}-g_{\mu\sigma}g_{\nu\rho}),\\
[g_{(2)}]^{\mu\nu\rho\sigma}&=\frac12(g^{\mu\rho}g^{\nu\sigma}-g^{\mu\sigma}g^{\nu\rho}).
\end{align}
Their contraction is the identity on antisymmetric index pairs:
\begin{align}
[g_{(2)}]_{\mu\nu\alpha\beta}[g_{(2)}]^{\alpha\beta\rho\sigma}&=\delta_{\mu\nu}^{\rho\sigma},\\
\delta_{\mu\nu}^{\rho\sigma}&\equiv\frac12\left(\delta_{\mu}^{\rho}\delta_{\nu}^{\sigma}-\delta_{\mu}^{\sigma}\delta_{\nu}^{\rho}\right).
\label{eq:antisymmetricPairIdentity}
\end{align}
Hence, for every antisymmetric tensor $X_{\rho\sigma}$,
\begin{equation}
\delta_{\mu\nu}^{\rho\sigma}X_{\rho\sigma}=X_{\mu\nu},
\qquad
[g_{(2)}]^{\mu\nu\rho\sigma}X_{\rho\sigma}=X^{\mu\nu}.
\label{eq:antisymmetricPairAction}
\end{equation}

The factors of one half belong to the ordered-pair Einstein summation convention. If we instead choose an explicit six-element basis of independent pairs $[\mu\nu]$, the corresponding ordinary $6\times6$ matrix entries absorb these combinatorial factors. Mixing the two conventions is the usual source of an erroneous factor of two; all traces in this paper use Eqs.~\eqref{eq:antisymmetricPairIdentity} and \eqref{eq:antisymmetricPairAction}.
Consequently, the trace of the curvature endomorphism is
\begin{align}
\Tr_{2}\cR
&=R_{\mu\nu\rho\sigma}[g_{(2)}]^{\rho\sigma\mu\nu}\\
&=\frac12R_{\mu\nu\rho\sigma}\left(g^{\rho\mu}g^{\sigma\nu}-g^{\rho\nu}g^{\sigma\mu}\right)\\
&=\frac12\left(R-(-R)\right)=R.
\label{eq:twoFormTraceDetailed}
\end{align}
In the second contraction, the inverse bivector metric first raises the antisymmetric pair:
\begin{equation}
R_{\mu\nu\rho\sigma}[g_{(2)}]^{\rho\sigma\alpha\beta}=R_{\mu\nu}{}^{\alpha\beta}.
\end{equation}

The remaining inverse bivector metric then implements the trace on the second pair, so
\begin{align}
\Tr_{2}\cR^{2}
&=R_{\mu\nu}{}^{\alpha\beta}R_{\alpha\beta\gamma\delta}[g_{(2)}]^{\gamma\delta\mu\nu}\\
&=R_{\mu\nu}{}^{\alpha\beta}R_{\alpha\beta}{}^{\mu\nu}\\
&=R_{\mu\nu\rho\sigma}R^{\mu\nu\rho\sigma}.
\label{eq:twoFormTraceSquaredDetailed}
\end{align}
No additional factor of $1/2$ appears: each bivector metric acts as the identity on an antisymmetric pair according to Eq.~\eqref{eq:antisymmetricPairAction}.

These identities reproduce the trace normalization used in the flattened-matrix construction of Ref.~\cite{Bianconi:2024aju}.

\section{Explicit scalar tidal calculation}
\label{app:tidal}

For $h_{\mu\nu}=\eta_{\mu\nu}\psi$,
\begin{equation}
h_{00}=-\psi,
\qquad h_{0i}=0,
\qquad h_{ij}=\delta_{ij}\psi.
\end{equation}
The linearized Riemann tensor gives
\begin{align}
R^{(1)}_{0i0j}&=\frac12\left(\partial_{i}\partial_{0}h_{0j}+\partial_{j}\partial_{0}h_{0i}-\partial_{i}\partial_{j}h_{00}-\partial_{0}^{2}h_{ij}
\right)\\
&=\frac12\partial_{i}\partial_{j}\psi-\frac12\delta_{ij}\partial_{0}^{2}\psi.
\end{align}
For Eq.~\eqref{eq:scalarplane},
\begin{equation}
\partial_{0}^{2}\psi=-\omega^{2}\psi,
\qquad
\partial_{z}^{2}\psi=-k^{2}\psi,
\qquad
\partial_{x}\psi=\partial_{y}\psi=0,
\end{equation}
which immediately gives Eqs.~\eqref{eq:breathing}--\eqref{eq:longitudinal}.

\bibliography{bibliography_GfE}

\end{document}